%% file: MAIN.tex
\documentclass[acmsmall,screen]{acmart}

\AtBeginDocument{%
  \providecommand\BibTeX{{%
    \normalfont B\kern-0.5em{\scshape i\kern-0.25em b}\kern-0.8em\TeX}}}

\setcopyright{rightsretained}
\setcctype{by-nc}
\acmYear{2025}
\acmJournal{CSUR}
\acmDOI{10.1145/3716845}
\acmVolume{57} 
\acmNumber{7} 
\acmMonth{2} 

\usepackage{graphicx}
\usepackage{array,multirow}
\usepackage{threeparttable}
\usepackage{enumitem}
\usepackage{makecell}
\usepackage{pifont}

\begin{document}

\title{Approximate Computing Survey, Part I: Terminology and Software \& Hardware Approximation Techniques}

\author{Vasileios Leon}
  \affiliation{%
  \institution{National Technical University of Athens}
  \department{School of Electrical and Computer Engineering}
  \country{Greece}}
\author{Muhammad Abdullah Hanif}
  \affiliation{%
  \institution{New York University Abu Dhabi}
  \department{Division of Engineering}
  \country{United Arab Emirates}}
\author{Giorgos Armeniakos}
  \affiliation{%
  \institution{National Technical University of Athens}
  \department{School of Electrical and Computer Engineering}
  \country{Greece}}
\author{Xun Jiao}
  \affiliation{%
  \institution{Villanova University}
  \department{Department of Electrical and Computer Engineering}
  \country{United States}}
\author{Muhammad Shafique}
  \affiliation{%
  \institution{New York University Abu Dhabi}
  \department{Division of Engineering}
  \country{United Arab Emirates}}
\author{Kiamal Pekmestzi}
  \affiliation{%
  \institution{National Technical University of Athens}
  \department{School of Electrical and Computer Engineering}
  \country{Greece}}
\author{Dimitrios Soudris}
  \affiliation{%
  \institution{National Technical University of Athens}
  \department{School of Electrical and Computer Engineering}
  \country{Greece}}

\renewcommand{\shortauthors}{Vasileios Leon et al.}
\renewcommand{\shorttitle}{Approximate Computing Survey, Part I: Terminology and Software \& Hardware Techniques}

\input{0_abstract}



\received{18 July 2023}
\received[Revised]{13 September 2024}
\received[Accepted]{4 February 2025}
\received[Published]{5 March 2025}

\maketitle

\newpage
\input{1_introduction}
\input{2_survey}
\input{3_terminology}
\input{4_software}
\input{5_hardware}
\input{6_comparison}
\input{7_conclusion}
\input{8_acknowledgement}

\vspace{30pt}

\bibliographystyle{ACM-Reference-Format}
\bibliography{REFs.bib}

\end{document}

%% file: 0_abstract.tex
\begin{abstract}
The rapid growth of demanding applications in domains applying multimedia processing and machine learning has marked a new era for edge and cloud computing. These applications involve massive data and compute-intensive tasks, and thus, typical computing paradigms in embedded systems and data centers are stressed to meet the worldwide demand for high performance. Concurrently, over the last 15 years, the semiconductor industry has established power efficiency as a first-class design concern. As a result, the community of computing systems is forced to find alternative design approaches to facilitate high-performance and power-efficient computing. Among the examined solutions, \emph{Approximate Computing} has attracted an ever-increasing interest, which has resulted in novel approximation techniques for all the layers of the traditional computing stack. More specifically, during the last decade, a plethora of approximation techniques in software (programs, frameworks, compilers, runtimes, languages), hardware (circuits, accelerators), and architectures (processors, memories) have been proposed in the literature. The current article is Part I of a comprehensive survey on Approximate Computing. It reviews its motivation, terminology and principles, as well it classifies the state-of-the-art software \& hardware approximation techniques, presents their technical details, and reports a comparative quantitative analysis.
\end{abstract}

%% file: 1_introduction.tex
\section{Introduction}
The proliferation of emerging technologies such as Artificial Intelligence (AI), Machine Learning (ML), Digital Signal Processing (DSP), big data analytics, cloud computing and Internet of Things (IoT) is driving the growing demand for computational power and storage requirements.
The International Data Corporation (IDC) reported that the global data sphere is expected to grow from 33 zettabytes (2018) to 175 zettabytes by 2025 with a Compound Annual Growth Rate (CAGR)  
of 61\%~\cite{forbes}, highlighting the pressing need for more efficient computing solutions.
This problem is intensified, especially when considering resource-restricted systems and/or battery-driven devices, such as smartphones and wearables~\cite{flexicore}.

Historically,
the industry of computing systems was driven for more than 40 years 
by two fundamental principles:
Moore's Law \cite{moore}
and 
Dennard's Law \cite{dennard}.
Today,
even though
the number of transistors integrated per area is still increasing (Moore's Law),
the supply voltage cannot be scaled 
according to Dennard's Law,
and thus,
the power density is increased. 
The end of Dennard's scaling 
combined with other factors
(e.g., the cooling technology
and the natural limits of silicon) 
led us to the ``Dark Silicon'' era \cite{dark2}.
In this era,
the power efficiency is a critical issue
for computing systems,
either they are placed at the edge (embedded systems)
or on the cloud (data centers). 
Concurrently, 
the compute-intensive workloads 
of novel AI/ML and DSP applications
challenge their deployment 
in terms of performance (speed). 
As a result, 
the industry of computing systems
is forced to find new 
design/computing approaches that 
will improve the power efficiency
while providing the desired performance.

\noindent
\textbf{I: Need for Low-Power/Energy Computing.}
With the continuous shrinking of the transistor size into deep nanometer regime, the power/energy consumption has become a critical issue and a top priority to consider in the design of computing systems.  
Actually, with the current trend, scientists have predicted that by the year 2040 computers will need more electricity than the world's  energy resources can generate, unless radical improvements are made in the computer design~\cite{electricity}. 
The ever-increasing deployment of IoT devices ~\cite{gartner, iotanal},  
the exploding ``Big Data'' from all kinds of sources 
(e.g., videos and images), and the growth of supporting cyberinfrastructure such as data centers, are all exaggerating this situation. 

This challenge is present in computing devices of all sizes --- from low-power edge devices to high-performance data centers. For example, for mobile/edge devices used intensively in IoT endpoints, low-power/energy computing must be achieved to increase the battery life. On the other hand, data centers must achieve low-power/energy to reduce the costs in electricity and cooling, and achieve reliable operations. In certain applications (e.g., autonomous driving), the high power consumption could lead to an increased temperature of the computing chips, which will adversely affect the reliability of the chips and cause severe consequences.  
Recently, the increasing workload (in terms of data size and computational demands) 
from the AI, ML, and DSP domains 
are all worsening the issue of power/energy consumption. 

\noindent
\textbf{II: Need for Accelerated Computing.} 
Many practical application domains, 
such as autonomous driving, robotics, and space, 
require real-time processing of data streams. However, the very nature of existing powerful algorithms pose significant challenges to the hardware implementations. 
A specific example is the recent massive deployment of AI/ML methods with millions of parameters, 
like in the case of Deep Neural Networks (DNNs). 
Typically speaking, the execution of DNN models requires a huge amount of computing operations, 
such as additions/multiplications and transformations, as well as intensive memory accesses, which may lead to a significant delay in processing data streams. This can cause compromised quality of results, especially in resource-constrained edge devices that have real-time latency requirements.

\noindent
\textbf{How Can Approximate Computing Help?}
As Dennard's Law expired in the mid-2000s
and Moore’s Law is declining,  
the transistor scaling is increasingly less effective in improving performance, energy efficiency, and robustness.
Therefore, alternative computing paradigms are urgently needed as we look to the future of the computing industry. 
\textbf{\textit{Approximate Computing (AxC)}} 
has recently arisen as a promising candidate for resolving this challenge due to its success in many compute-intensive applications
(e.g., image processing, object classification, and bio-signal analysis). 
Such applications show an inherent error tolerance, i.e., they do not require completely accurate computations for delivering acceptable output quality. For example, in image processing, a few pixel drops do not affect how images are perceived by human eyes; AI/ML may not need precise model parameters to get accurate results in classification and detection; communication systems are resilient against occasional noise. 
This paves the way for new optimization opportunities. 
By introducing a new design dimension -- ``accuracy'' -- to the overall design optimization, it is possible to trade off accuracy and lead to \textit{less power used, less time consumed, and fewer computing resources required}. This leads to the promising novel design paradigm of Approximate Computing.  
Some examples of accuracy/quality metrics are peak signal-to-noise ratio (multimedia applications), relative difference (numerical analysis), and classification accuracy (machine learning). 

The idea of leveraging imprecise computation for improved design dates back several decades in real-time system scheduling, where imprecise computation was used to enhance its dependability~\cite{liu1994use}. Another related research field is Fault Tolerance, which seeks to continue to provide the required functionality despite occasional failures by hiding the errors~\cite{pierce2014failure}. 
Compared to this field, 
Approximate Computing ``intentionally'' seeks to design imperfect hardware and software systems (\emph{induce  errors}) for improved performance and/or power/area efficiency. 
Specifically, researchers have built approximate integrated circuits, software programs, and architectures that
outperform their conventional ``accurate'' counterparts in terms of resources (power, area, and/or performance).  
Approximate Computing has achieved tremendous success in many application domains 
and targets among other 
image processing~\cite{2020_Adams_IEEEtc, 2017_Akbari_IEEEtvlsi}, computer vision~\cite{2015_Achour_OOPSLA, 2020_Baharvand_IEEEtetc}, computer graphics~\cite{2010_Baek_PLDI, 2020_Baharvand_IEEEtetc}, machine learning~\cite{2016_Anderson_ICDE, 2010_Baek_PLDI}, signal processing~\cite{2005_Alvarez_IEEEtc, 2010_Baek_PLDI}, financial analysis~\cite{2015_Achour_OOPSLA, 2010_Baek_PLDI}, database search~\cite{2013_Agarwal_EuroSys, 2010_Baek_PLDI}, and scientific computing~\cite{2011_Ansel_CGO, 2015_Boston_OOPSLA}. 
The error resilience of such application domains 
and the relaxed constraints for the quality of the results 
constitute Approximate Computing as an applicable design paradigm.
They originate from:
\begin{enumerate}[itemsep=0.3pt,topsep=5pt]
    \item The user's intention to accept inaccuracies and results of lower quality.
    \item The limited human perception, e.g., in multimedia applications.
    \item The lack of perfect/golden results for validation, e.g., in data mining applications.
    \item The lack of a unique answer/solution, e.g., in machine learning applications.
    \item The application's self-healing property, i.e., its capability to absorb/compensate errors by default.   
    \item The application's inherent approximate nature, e.g., in probabilistic calculations, iterative algorithms, and learning systems.
    \item The application's analog/noisy real-world input data, e.g., in multimedia/signal processing.
\end{enumerate}
The prominent outcome of approximate systems,
in parallel with the ever-increasing demand for efficient and sustainable computing, has attracted a vast research interest.
In this context, 
approximation techniques are applied 
at different design layers,
i.e., from software/programs to hardware/circuits.
Motivated by the benefits of Approximate Computing,
as well as the great momentum it has gained over the last years, 
we conduct a survey that is presented in two parts.
The goal is 
to cover the entire spectrum of Approximate Computing.
The contribution and the content of the two-part survey are analyzed in 
Section \ref{sec:surv}. 

%% file: 2_survey.tex
\section{Approximate Computing Survey}\label{sec:surv}

\begin{figure}[!t]
    \centering
    \includegraphics[width=1\textwidth]{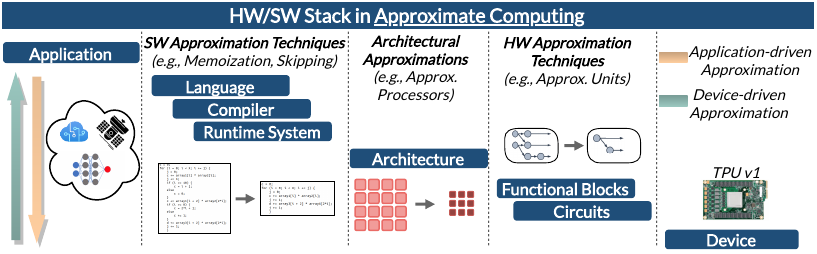}
    \vspace{-19pt}
    \caption{The Approximate Computing stack: approximation techniques in the design abstraction layers.}
    \label{fg_ac}
\end{figure}

\textbf{Scope and Contribution:}
The literature includes surveys on Approximate Computing
that target specific areas, e.g., arithmetic circuits \cite{2020_Jiang_IEEE} and logic synthesis \cite{2020_Scarabottolo_IEEE},
or focus on a single approximation technique,
e.g., precision scaling \cite{2020_Cherubin_ACMsrv}.
In~\cite{swagath:2020} and~\cite{csur:arme},  
a thorough analysis of software and hardware approximation techniques, respectively, is provided, but they are both laser-focused on DNN applications. 
Similarly, the survey of \cite{edge:survey} mainly reviews the impact of AxC on edge computing and none of the~\cite{2020_Jiang_IEEE, 2020_Scarabottolo_IEEE, 2020_Cherubin_ACMsrv, swagath:2020, csur:arme, edge:survey} surveys cover approximations from system level down to circuit level.
\emph{In contrast, the proposed two-part survey covers the entire computing stack, 
reviewing and classifying all the state-of-the-art approximation techniques from the software, hardware, and architecture layers for a wide range of application domains.}
Table~\ref{tb_srv} reports a list of surveys that can be classified along with the proposed one, showing a qualitative comparison and key aspects that characterize each work.
In the current work, 
the AxC stack (pyramid of design layers) is formed as shown in Fig.~\ref{fg_ac}, and design techniques \& approaches from all the layers are analyzed.  
In fact, this is the traditional computing stack with the addition of various kind of approximation techniques 
across the design layers
(from the application and software
down to the hardware and device).
It is worth mentioning that the proposed two-part survey is the first to report a quantitative analysis with numerical assessments at this extent. 

\input{TBs/tb1_survey_comparison}

More explicitly, 
the current survey constitutes a comprehensive and detailed guide that provides a step-by-step explanation of key concepts, techniques, and applications of the AxC paradigm.
The reader will have a complete view on AxC principles and works that implement and evaluate software, hardware, application-specific and architectural approximations.
This survey also acts as a tutorial on the state-of-the-art approximation techniques.
The main objectives and contributions of the survey are: 
1) to attribute definitions in key AxC aspects and explain the main terminology, 
2) to analyze the state-of-the-art works, identify approximation categories and cluster the reviewed works with respect to the approximation type/approach,  
3) to survey application domains of AxC including the impact of approximations on them, 
and 4) to identify and discuss open challenges and future directions as a step towards the realization of approximate applications. 

\begin{figure}[t]
    \centering
    \includegraphics[width=1\textwidth]{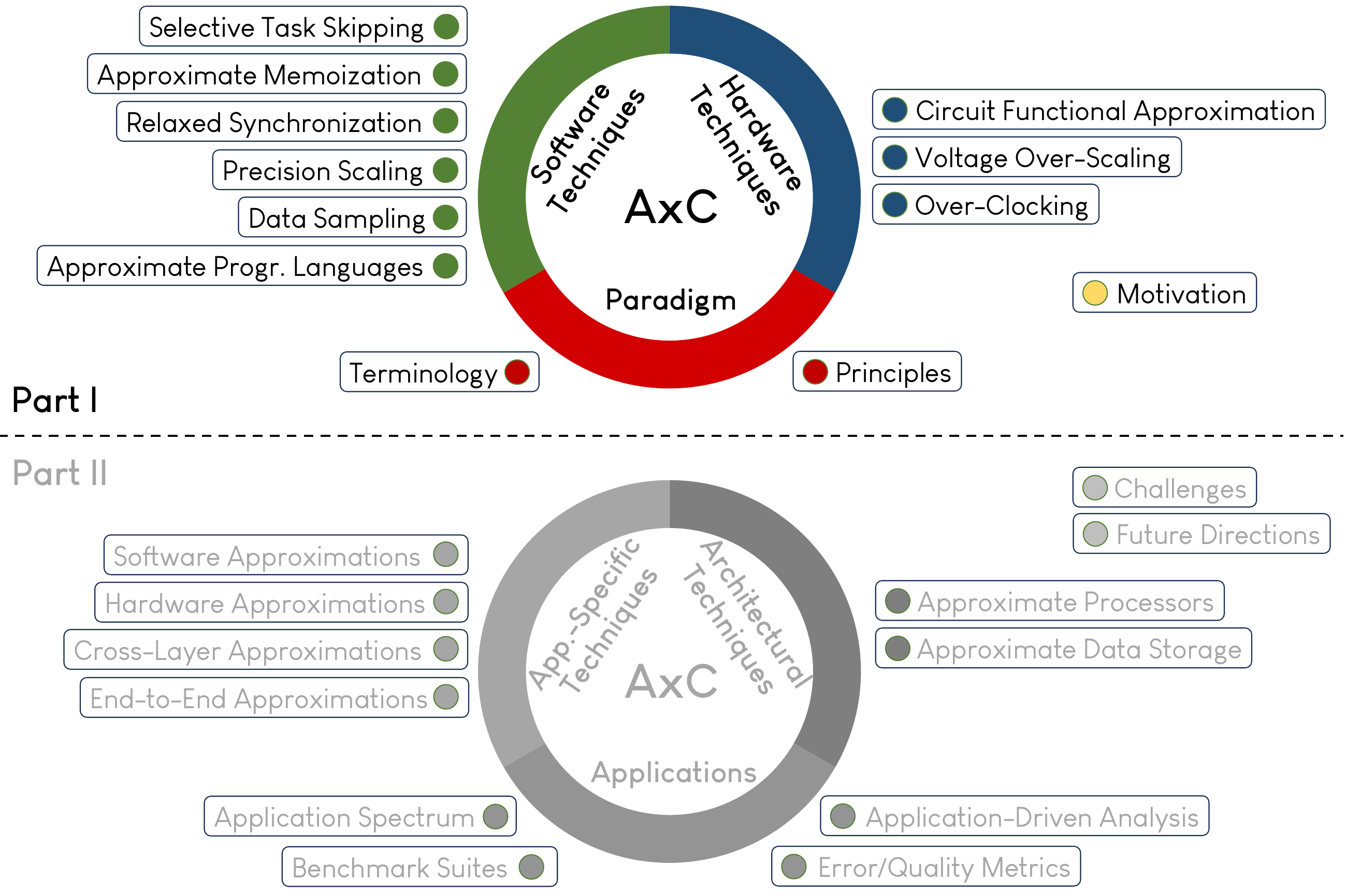}
    \vspace{-19pt}
    \caption{Organization of the proposed two-part survey on Approximate Computing.}
    \label{fg_struct}
\end{figure}

\noindent\textbf{Organization:}
As shown in Fig.~\ref{fg_struct}, 
the proposed survey is divided into two parts,
which constitute standalone manuscripts focusing on different aspects/areas of Approximate Computing:

\begin{description}[leftmargin=35pt]
\item[\hspace{3.8pt}Part I:] It is presented in the current paper, and it introduces the AxC paradigm
(terminology and principles)
and reviews software \& hardware approximation techniques. 
\item[Part II:] It is presented in \cite{mysurvey_pt2}, and it reviews application-specific \& architectural approximation techniques and introduces the AxC applications  (domains, quality metrics, benchmarks). 
\end{description}

The remainder of the article (Part I of the survey)
is organized as follows. 
Section~\ref{sec:paradigm} provides fundamental concepts of AxC, while the next two sections (Sections~\ref{sec:sw_lvl}-\ref{sec:hw_lvl}) review and classify  
software- and hardware-level
works,
respectively. 
Section~\ref{sec:comp} performs a quantitative analysis for all the reviewed approximation techniques.  
Finally, Section~\ref{sec:conc} concludes the survey. 


%% file: TBs/tb1_survey_comparison.tex
\begin{table*}[!t]
\renewcommand{\arraystretch}{1.1}
\setlength{\tabcolsep}{3.4pt}
\centering
\footnotesize
\caption{Qualitative comparison of Approximate Computing surveys on the entire computing stack.}
\vspace{-8pt}
\begin{threeparttable}
\begin{tabular}{c@{\hspace{3.7pt}}cccccccccccccccc}
\hline
\multicolumn{2}{c|}{\rotatebox{45}{\textbf{\hspace{6pt}AxC Survey}}} &
  \multicolumn{1}{c|}{\rotatebox{80}{\textbf{Year Coverage}}} &
  \multicolumn{1}{c|}{\rotatebox{80}{\textbf{Pages \#}}} &
  \multicolumn{1}{c|}{\rotatebox{80}{\textbf{References \#}}} &
  \multicolumn{1}{c|}{\rotatebox{80}{\textbf{SW Tech.}}} &
  \multicolumn{1}{c|}{\rotatebox{80}{\textbf{HW Tech.}}} &
  \multicolumn{1}{c|}{\rotatebox{80}{\textbf{Arch. Approx.}}} &
  \multicolumn{1}{c|}{\rotatebox{80}{\textbf{AI/ML}}} &
  \multicolumn{1}{c|}{\rotatebox{80}{\textbf{Memories}}} &
  \multicolumn{1}{c|}{\rotatebox{80}{\textbf{\begin{tabular}[c]{@{}c@{}}Frameworks\\[-3pt] \& Tools\end{tabular}}}} &
  \multicolumn{1}{c|}{\rotatebox{80}{\textbf{Metrics}}} &
  \multicolumn{1}{c|}{\rotatebox{80}{\textbf{Benchmarks}}} &
  \multicolumn{1}{c|}{\rotatebox{80}{\textbf{\begin{tabular}[c]{@{}c@{}}CPU/FPGA/\\[-3pt] GPU/ASIC\end{tabular}}}} &
  \multicolumn{1}{c|}{\rotatebox{80}{\textbf{Terminology}}} 
  &
  \multicolumn{1}{c|}{\rotatebox{80}{\textbf{Challenges}}} 
  &
  \multicolumn{1}{c|}{\rotatebox{80}{\textbf{Quant. Analysis\tnote{2}\phantom{b}}}}
  \\ \hline \hline

\multicolumn{2}{c}{\cite{2013_Han_ETS}}            & 2013 & 6  & 65  & \ding{51}  & \ding{51}  & $\approx$ & $\approx$ & \ding{55}  & \ding{55}  & \ding{51} & \ding{55} & $\approx$ & \ding{55}  & \ding{55} & $\approx$ \\ \hline
  
\multicolumn{2}{c}{\cite{2016_Mittal_ACMsrv}}     & 2015 & 33\tnote{1}  & 84  & \ding{51}  & $\approx$ & \ding{51}  & $\approx$ & \ding{51}  & \ding{51}  & \ding{51} & \ding{51} & \ding{51}  & \ding{55}  & \ding{51} & \ding{55} \\ \hline

\multicolumn{2}{c}{\cite{2016_Xu_IEEEdt}}         & 2015 & 15 & 59  & \ding{51}  & \ding{51}  & \ding{51}  & \ding{55}  & $\approx$ & \ding{55}  & \ding{55} & \ding{55} & $\approx$ & \ding{55}  & \ding{55} & \ding{55} \\ \hline

\multicolumn{2}{c}{\cite{2015_Venkataramani_DAC}} & 2015 & 6  & 54  & \ding{51}  & \ding{51}  & \ding{51}  & \ding{55}  & \ding{55}  & \ding{51}  & \ding{55} & \ding{55} & $\approx$ & \ding{55}  & \ding{55} & \ding{55} \\ \hline

\multicolumn{2}{c}{\cite{2016_Shafique_DAC}}      & 2016 & 6  & 47  & \ding{51}  & \ding{51}  & \ding{51}  & \ding{55}  & \ding{55}  & $\approx$ & \ding{55} & \ding{55} & $\approx$ & \ding{55}  & \ding{55} & $\approx$ \\ \hline

\multicolumn{2}{c}{\cite{2018_Moreau_IEEEesl}}    & 2017 & 4  & 40  & \ding{51}  & \ding{51}  & \ding{55}  & $\approx$ & $\approx$ & \ding{55}  & \ding{55} & \ding{55} & $\approx$ & \ding{55}  & \ding{55} & \ding{55} \\ \hline

\multicolumn{2}{c}{\cite{2018_Moreno_LATS}}       & 2017 & 6  & 72  & \ding{51}  & \ding{51}  & \ding{51}  & $\approx$ & $\approx$ & $\approx$ & \ding{55} & \ding{51} & $\approx$ & \ding{55}  & \ding{51} & \ding{55} \\ \hline 

\multicolumn{2}{c}{\cite{2020_Stanley_ACMsrv}}   & 2020 & 39\tnote{1}  & 235 & $\approx$ & \ding{51}  & \ding{51}  & \ding{55}  & \ding{51}  & $\approx$ & \ding{55} & \ding{55} & $\approx$ & \ding{55}  & \ding{51} & \ding{55} \\ \hline \hline

\multirow{2}{*}{This work} & Pt. 1       & 2023 & 36\tnote{1}  & 222 & \ding{51}  & \ding{51}  &  &   &   & \ding{51}  &  &  & \ding{51} & \ding{51}  &  & \ding{51}\\  

& Pt. 2       & 2023 & 36\tnote{1}  & 301 & \ding{51}  & \ding{51}  & \ding{51} & \ding{51}  & \ding{51}  &   & \ding{51} & \ding{51}  & \ding{51} &   & \ding{51} & \ding{51} \\ \hline \hline
\end{tabular}
\begin{tablenotes}
\scriptsize 
    \item[1] Single-column pages. 
    \item[2] Quantitative analysis involving count of works, frequencies and numerical assessments.
\end{tablenotes}
\end{threeparttable}
\label{tb_srv}
\end{table*}

%% file: 3_terminology.tex
\section{The Approximate Computing Paradigm}\label{sec:paradigm}

\subsection{Terminology of Approximate Computing}

Even though approximate computations have been examined 
since the 1960s
(e.g., Mitchell's logarithmic-based multiplication/division \cite{1962_Mitchell}),
the first systematic efforts to 
define the Approximate Computing paradigm
started in the late 2000s.  
Various terms have been used
in the literature 
to describe strategies 
for delivering approximate architectures, programs, and circuits.
Approximate Computing is synonymous or overlaps with these terms.
Chakradhar et al. \cite{2010_Chakradhar_DAC}
define \textbf{\emph{Best-Effort Computing}} as
\emph{``the approach of designing software/hardware computing systems
with reduced workload, improved parallelization and/or approximate components
towards enhanced efficiency and scalability''}.
The term \textbf{\emph{Relaxed Programming}} is introduced by
Carbin et al. \cite{2012_Carbin_PLDI}
to express 
\emph{``the transformation of programs with approximation methods and relaxed semantics
to enable greater flexibility in their execution''}.
Chippa et al. \cite{2014_Chippa_IEEEtvlsi}
use the term \textbf{\emph{Scalable Effort Design}} 
for
\emph{``the systematic approach that embodies the notion of scalable effort into the design process at different levels of abstraction,
involving mechanisms to vary the computational effort and control knobs to achieve the best possible trade-off between energy efficiency and quality of results''}.

According to Mittal \cite{2016_Mittal_ACMsrv},
\emph{``Approximate Computing exploits the gap between
the accuracy required by the applications/users and that provided by the computing system
to achieve diverse optimizations''}.
Han and Orshansky \cite{2013_Han_ETS}
distinguish Approximate Computing
from Probabilistic/Stochastic Computing,
stating that 
\emph{``it does not involve assumptions on the stochastic
nature of the underlying processes implementing the system
and employs deterministic designs for producing inaccurate results''}.
Another interesting point of view 
is expressed by Sampson \cite{2015_Sampson_phd},
who claims that 
Approximate Computing 
is based on
\emph{``the idea that we are hindering the efficiency of the computer systems
by demanding too much accuracy from them''}.
The current article attributes the following definition:\\

\noindent
\textbf{Approximate Computing:} 
\emph{It constitutes a radical paradigm shift in the design and development of computing systems, circuits and programs, which is built on top of the error-resilient nature of various application domains and relies on disciplined methods to intentionally insert errors for providing valuable resource gains in exchange for tunable accuracy loss}.\\

\input{TBs/tb2_terminology}

Table \ref{tb_terms} describes the most frequently used terms in Approximate Computing. 
The term \emph{error} is used to 
indicate that the output result
is different from 
the accurate result (produced with conventional computing). 
Error is distinguished from \emph{fault}, 
which refers to an unexpected condition
(e.g., stuck-at-logic in circuits, bit-flips in memories, faults in operating systems) 
that causes the system to unintentionally output erroneous results. 
Another significant term is
\emph{accuracy},
which is defined as 
the distance between the approximate and the accurate result
and is measured using application-specific and general-computing error metrics.
Accuracy is distinguished from 
\emph{precision},
which expresses the differentiation between nearby discrete values
and does not refer to errors of Approximate Computing but to quantization noise
(inserted by the real-to-digital value mapping). 
Moreover,
in Approximate Computing,  
the term \emph{Quality-of-Service (QoS)} is used 
to describe the overall quality of the results regarding accuracy and errors.

\subsection{Principles of Approximate Computing}

To enable and realize significant efficiency gains through approximations, the design of approximate systems should be guided using the following steps/principles: 

-- \emph{Application Analysis}: The quality requirements and metrics vary across applications. Therefore, it is essential to analyze the application in detail to identify the acceptable QoS and specify the error metrics that can truly quantify the output quality for evaluation and comparison. 

-- \emph{Workload Analysis}: Not all tasks/computations in an application can be approximated. Therefore, it is important to identify the non-critical tasks/computations (to be approximated) and isolate them from the critical ones. This is essential to enable disproportionate benefits, i.e., significant improvements in efficiency for a negligible loss in quality. 

-- \emph{Development of Approximation Methodology}: To achieve ultra-high efficiency gains while ensuring that acceptable quality is maintained, approximations are required to be introduced systematically in the system. Moreover, the sources of disproportionate benefits are distributed across different layers of the computing stack. Therefore, to achieve a superior quality--efficiency trade-off, the development of sophisticated methodologies that can exploit the cross-layer knowledge of the system and deploy approximations systematically across various layers and sub-systems of the given system becomes an important step. 

-- \emph{Development of Error Models}: Error estimation is vital for comparing different approximations. However, empirical evaluation of approximate implementations is time-consuming and costly (especially when inducing approximations at multiple design layers or at low-level hardware implementations).  
Therefore, it is essential to build models that can emulate the errors and examine the output quality of the system when approximated. 

-- \emph{Design Space Exploration}: Typically, various types of approximations can be deployed in a system, where each sub-system/system module may have a completely different set of approximations that lead to disproportionate benefits. Therefore, design space exploration is performed to examine different approximation configurations, evaluate the approximation space and make decisions regarding the final approximate implementation that yields significant efficiency gains while meeting user-defined quality and performance constraints. These exploration methodologies are usually supported by error and performance models to efficiently search the approximation space.  

-- \emph{Error Analysis}: Input distribution can have a profound impact on the resilience of the approximated system. Therefore, it is important to study the errors for different input distributions using appropriate error/QoS metrics. 

-- \emph{Quantification of Results}: This is performed to prove the resilience of the application and ensure that constraints about QoS and/or resources are met.

%% file: TBs/tb2_terminology.tex
\begin{table*}[!t]
\renewcommand{\arraystretch}{1.1}
\centering
\footnotesize
\setlength{\tabcolsep}{3.8pt}
\caption{Terminology of Approximate Computing.}
\vspace{-8pt}
\begin{tabular}{l| m{9.5cm}}
\hline  
\makecell[c]{\textbf{Term}} & \makecell[c]{\textbf{Description}}\\
\hline
\hline
\emph{Error-Resilient Application} & The application that allows computation errors and accepts results of lower quality. \\ \hline
\emph{Quality of Service}                   & The quality of the results in terms of errors and accuracy. \\ \hline
\emph{Accuracy Constraint}                  & The quality requirements that the results need to satisfy. \\ \hline
\emph{Error Bound/Threshold}                & The maximum error allowed in the results. \\ \hline
\emph{Golden Result}                        & The result that is obtained from the original accurate computations. \\ \hline
\emph{Acceptable Result}                    & The result that satisfies the accuracy constraints and error bounds of the application. \\ \hline
\emph{Variable Accuracy}                    & The capability of providing different levels of accuracy. \\ \hline
\emph{Non-Critical Task/Computation}        & The task/computation that can be safely approximated due to its small impact on the quality of the output results. \\ \hline
\emph{Error Analysis}                       & The study involving metrics, mathematics and simulations to examine the range, frequency, scaling, and/or propagation of the errors. \\ \hline
\emph{Approximation Technique/Method}       & The systematic and disciplined approach to insert computation errors in exchange for gains in power, energy, area, latency, and/or throughput. \\ \hline 
\emph{Approximation Degree/Strength}        & The aggressiveness of the approximation technique in terms of errors inserted and tasks/computations approximated. \\ \hline
\emph{Approximation Configuration}          & An instance of the parameters and settings of the approximation technique. \\ \hline
\emph{Frozen Approximation}                 & The approximation that is fixed and cannot be re-configured at a different degree. \\ \hline
\emph{Dynamic Approximation Tuning}         & The capability of adjusting the approximation degree at runtime to satisfy the desired error constraints. \\ \hline
\emph{Cross-Layer Approximation}            & The approximation that is applied at multiple design abstraction layers (software, hardware, architecture). \\ \hline
\emph{Heterogeneous Approximation}          & The approximation that applies concurrently multiple configurations of different degree within the same system. \\ \hline
\emph{Application-Driven Approximation}           & The approximation that is applied with respect to the error resilience and sensitivity of the targeted application. \\ \hline
\emph{Device-Driven Approximation}           & The approximation that is applied with respect to the targeted device/technology (e.g., CPU, GPU, ASIC, FPGA). \\ \hline
\emph{Approximate Space Exploration}        & The study involving error analysis and resource gain quantification to examine trade-offs and select the most suitable approximation techniques/configurations. \\ \hline
\emph{Approximation Localization}           & The systematic approach to locate the tasks/computations and design regions that are offered for approximation. \\ \hline
\emph{Error Modeling}                       & The process of emulating the errors inserted by the approximations. \\ \hline
\emph{Error Prediction}                     & The process of predicting errors before computing the final result. \\ \hline
\emph{Error Detection}                      & The process of identifying an error occurrence. \\ \hline
\emph{Error Compensation}                   & The process of modifying the erroneous result to reduce the error. \\ \hline
\emph{Error Correction}                     & The process of replacing the erroneous result with the accurate one.  \\ \hline\hline
\end{tabular}
\label{tb_terms}
\end{table*}

%% file: 4_software.tex
\section{Software Approximation Techniques}\label{sec:sw_lvl}

\begin{figure}[!t]
    \centering
    \includegraphics[width=0.85\textwidth]{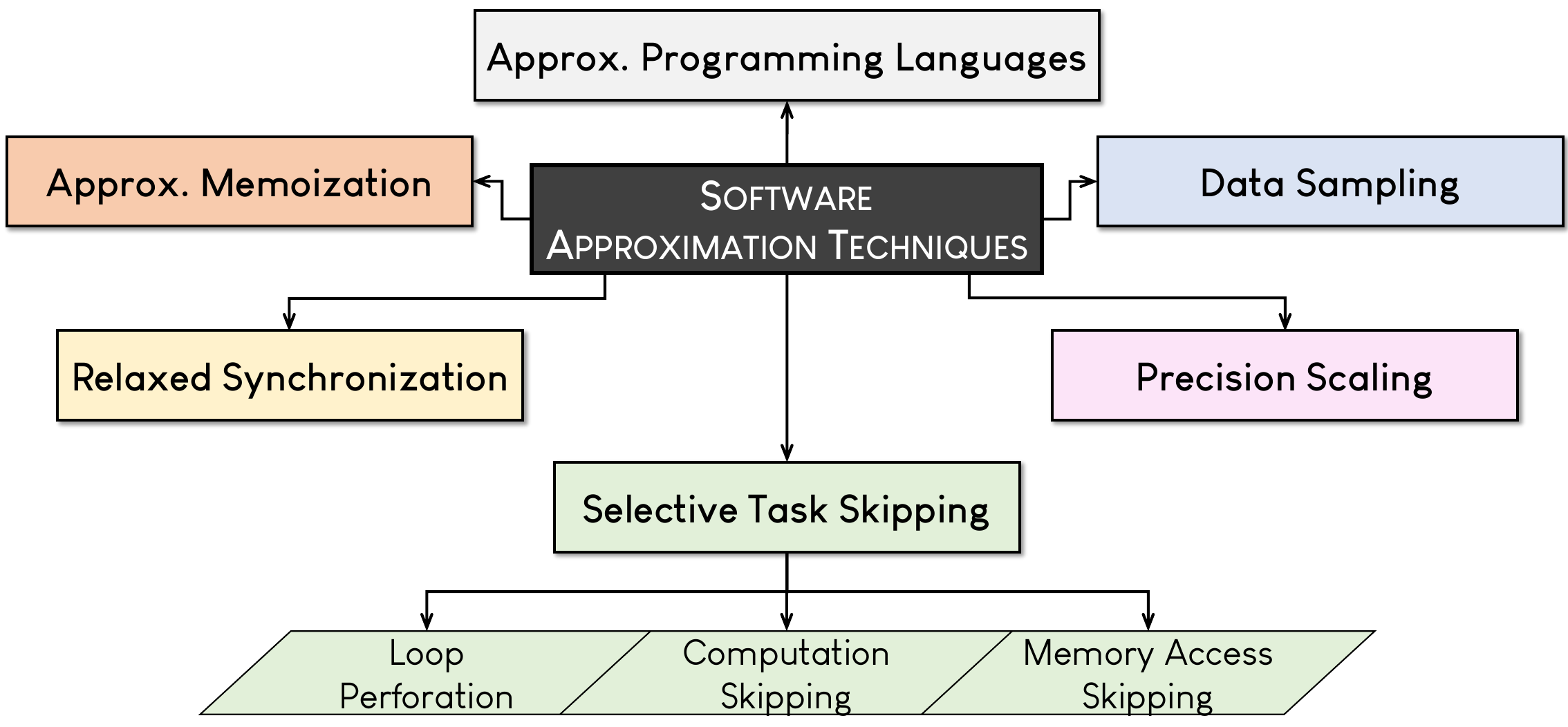}
    \vspace{-6pt}
    \caption{Classification of SW approximation techniques in 6 classes:
    \emph{Selective Task Skipping}, \emph{Approximate Memoization}, \emph{Relaxed Synchronization}, \emph{Precision Scaling}, \emph{Data Sampling}, and 
    \emph{Approximate Programming Languages}.}
    \label{fg_sw}
\end{figure}

This section classifies and presents approximation techniques
that are applied at software level,
i.e., the higher level of the design abstraction hierarchy.
The goal of software Approximate Computing
is to improve the execution time of the program
and/or the energy consumption of the system.
The techniques of the literature,
illustrated in Fig.~\ref{fg_sw},
can be categorized into five classes:
(i) \emph{Selective Task Skipping},
(ii) \emph{Approximate Memoization},
(iii) \emph{Relaxed Synchronization},
(iv) \emph{Precision Scaling},
(v) \emph{Data Sampling},
and (vi) \emph{Approximate Programming Languages}.
Typical software approximation techniques
integrate some of the following features:
approximation libraries/frameworks,
compiler extensions,
accuracy tuning tools,
runtime systems, 
and language annotations. 
Moreover, 
numerous of these techniques
allow the programmer to 
specify QoS constraints,
provide approximate code variants,
and mark the program regions/tasks for approximation. 

The remainder of this section
reports representative state-of-the-art works
for software approximation techniques. 
The references of these works are summarized in Table \ref{tb_sw}.
The literature also includes software approximation frameworks,
such as ACCEPT \cite{2015_Sampson_UOW} and OPPROX \cite{2017_Mitra_CGO},
which apply multiple state-of-the-art approximation techniques.

\subsection{Selective Task Skipping}

\subsubsection{Loop Perforation}
The loop perforation technique aims at skipping some of the loop iterations in a software program to provide performance/energy gains in exchange for QoS loss.
Subsequently, 
several relevant works \cite{2009_Hoffmann_MIT, 2010_Misailovic_ICSE, 2011_Sidiroglou_FCE, 2015_Shi_IEEEcal, 2017_Omar_ICCD, 2018_Li_ICS, 2020_Baharvand_IEEEtetc, 2015_Tan_ASP-DAC, 2018_Kanduri_DAC},  
which involve design space exploration on loop perforation
with programming frameworks and profiling tools,  
are presented. 

Starting with one of the first state-of-the-art works, 
the SpeedPress compiler \cite{2009_Hoffmann_MIT}
supports a wide range of loop perforation types,
i.e., modulo, truncation, and randomized.
It takes as input the original source code,
a set of representative inputs,
as well as a programmer-defined QoS acceptability model,
and outputs a loop perforated binary.
In the same context, 
Misailovic et al. \cite{2010_Misailovic_ICSE}
propose a QoS profiler
to identify computations that can be approximated via
loop perforation. 
The proposed profiling tool
searches the space of loop perforation
and generates results
for multiple perforation configurations.
In \cite{2011_Sidiroglou_FCE}, 
the same authors propose a methodology
to exclude critical loops, 
i.e., whose skipping results in unacceptable QoS,
and perform exhaustive and greedy design space explorations to 
find the Pareto-optimal perforation configurations
for a given QoS constraint. 

\input{TBs/tb3_sw_classification}

In \cite{2015_Shi_IEEEcal},
the authors propose an architecture 
that employs a profiler 
to identify non-critical loops towards their perforation. 
To protect code segments 
that can be affected by the perforated loops,
the architecture is equipped with
HaRE,
i.e., a hardware resilience mechanism.
Another interesting work is
GraphTune \cite{2017_Omar_ICCD},
which is an input-aware loop perforation scheme
for graph algorithms. 
This approach 
analyzes the input dependence of graphs
to build
a predictive model that finds
near-optimal perforation configurations
for a given accuracy constraint. 
Li et al. \cite{2018_Li_ICS}
propose a compiling \& profiling system,
called Sculptor,
to improve the conventional loop perforation,
which skips a static subset of iterations. 
More specifically,
Sculptor dynamically skips a subset of the loop instructions (and not entire iterations)
that do not affect the output accuracy. 
More recently, 
the authors of \cite{2020_Baharvand_IEEEtetc} develop LEXACT,
which is a tool for identifying non-critical code segments
and monitoring the QoS of the program.
LEXACT searches the loop perforation space, 
trying to find perforation configurations that satisfy pre-defined metrics. 

The loop perforation technique has been also used 
in approximation frameworks for heterogeneous multi-core systems combining various approximation mechanisms.  
Tan et al. \cite{2015_Tan_ASP-DAC}
propose a task scheduling algorithm that 
employs multiple approximate versions of the tasks
with loops perforated. 
Kanduri et al. \cite{2018_Kanduri_DAC}
target applications whose main computations are continuously repeated 
and tune the loop perforation at runtime. 

\subsubsection{Computation Skipping}
This technique 
omits the execution of blocks of codes
according to 
the acceptable QoS loss,
programmer-defined constraints,
and/or runtime predictions regarding the output accuracy
\cite{2009_Meng_IPDPS, 2010_Byna_GPGPU, 2015_Raha_DATE, 2015_Vassiliadis_PPoPP, 2016_Vassiliadis_CGO, 2006_Rinard_ICS, 2007_Rinard_OOPSLA, 2017_Lin_ISCAS, 2018_Akhlaghi_ISCA}. 
Compared to loop perforation,
these techniques do not focus only on skipping loop iterations,
but also skip higher-level computations/tasks 
e.g., an entire convolution operation. 
Most of the state-of-the-art works perform application-specific computation skipping.

Meng et al. \cite{2009_Meng_IPDPS}
introduce a parallel template
to develop 
approximate programs 
for iterative-convergence recognition \& mining algorithms.
The proposed programming template
provides several strategies (implemented as libraries)
for task dropping,
such as convergence-based computation pruning,
computation grouping in stages,
and early termination of iterations. 
Another interesting work involving application-specific computation skipping is presented in \cite{2010_Byna_GPGPU}. 
The authors of this work study 
the error tolerance of
the supervised semantic indexing algorithm
to make approximation decisions.
Regarding their task dropping approach,
they choose to omit 
the processing of common words (e.g., ``the'', ``and'') after the initial iterations, 
as these 
computations have negligible impact on the output accuracy. 

The authors of
\cite{2015_Raha_DATE}
propose two techniques
to find computations with low impact on the 
QoS of the
Reduce-and-Rank
computation pattern,
targeting to approximate or skip them completely. 
To identify these computations,  
the first technique 
uses intermediate reduction results and ranks,
while the second one 
is based on the spatial or temporal correlation of the input data
(e.g., adjacent image pixels or successive video frames).
Similarly to the other state-of-the-art works,
Vassiliadis et al. \cite{2015_Vassiliadis_PPoPP, 2016_Vassiliadis_CGO}
propose a programming environment 
that skips (or approximates) computations according to programmer-defined QoS constraints.
More specifically, 
the programmer 
expresses the significance of the tasks using pragmas directives,
optionally provides approximate variants of tasks,
and specifies the desired task percentage to be executed accurately. 
Based on these constraints,
the proposed system
makes decisions at runtime  
regarding the approximation/skipping
of the less significant tasks. 

Rinard \cite{2006_Rinard_ICS} builds probabilistic distortion models 
based on linear regression
to study the impact of computation skipping on the output accuracy. 
In particular,
the programmer partitions the computations into tasks, 
which are then marked as 
``critical'' or ``skippable''
through random skip executions. 
The probabilistic models estimate the output distortion as a function of the skip rates of the skippable tasks.
This approach is also applied in parallel programs \cite{2007_Rinard_OOPSLA}, 
where probabilistic distortion models are employed 
to tune the early phase termination at barrier synchronization points,
targeting to keep all the parallel cores busy.

Significant research has been also conducted 
on skipping the computations of 
Convolutional Neural Networks (CNNs).
Lin et al. \cite{2017_Lin_ISCAS}
introduce PredictiveNet
to predict the sparse outputs of the nonlinear layers 
and 
skip a large subset of convolutions at runtime. 
The proposed technique,
which does not require 
any modification in the original CNN structure,
examines the most-significant part of the convolution
to predict if the nonlinear layer output is zero,
and then decides whether
to skip 
the remaining least-significant part computations
or not. 
In the same context, 
Akhlaghi et al. \cite{2018_Akhlaghi_ISCA}
propose SnaPEA,
exploiting the convolution--activation algorithmic chain in CNNs (activation takes as input the convolution result and outputs zero if the latter is negative).
This technique
early predicts 
negative convolution results 
based on static re-ordering of the weights 
and monitoring of the partial sums' sign bit,
in order to skip the rest computations.

\subsubsection{Memory Access Skipping}
Another approach 
to improve the execution time and energy consumption
at the software level
is the memory access skipping.
Such techniques \cite{2014_Miguel_MICRO, 2016_Yazdanbakhsh_ACMtaco, 2018_Kislal_ELSEclss, 2013_Samadi_MICRO, 2019_Karakoy_ACMmacs, 2015_Zhang_DATE} aim at 
avoiding high-latency memory operations,
while they 
inherently reduce the number of computations. 

Miguel et al. \cite{2014_Miguel_MICRO} exploit the approximate data locality 
to skip the required memory accesses due to L1 cache miss.
They employ a load value approximator,
which learns value patterns 
using a global history buffer and an approximator table, 
to estimate the memory data values.
RFVP \cite{2016_Yazdanbakhsh_ACMtaco}
uses value prediction instead of memory accessing.
When selected load operations miss in the cache memory,
RFVP predicts the requested vales
without checking for misprediction
or recovering the values.
Thus,
timing overheads from pipeline flushes
and re-executions
are avoided.
Furthermore,
a tunable rate of cache misses
is dropped
after the value prediction
to eliminate 
long memory stalls. 
Similarly,
the authors of \cite{2018_Kislal_ELSEclss}
propose a framework 
that skips costly last-level cache misses 
according to a programmer-defined error constraint
and an heuristic predicting skipped data. 

To improve the performance of CUDA kernels on GPUs, 
Samadi et al. \cite{2013_Samadi_MICRO}
propose a runtime approximation framework,
called SAGE,
which focuses on optimizing the memory operations among other functionalities.
The approximations lie in
skipping selective atomic operations
(used by kernels to write shared variables) 
to avoid conflicts leading to performance decrease.
Furthermore,
SAGE reduces the number of memory accesses
by packing the read-only input arrays, 
and thus, allowing to access more data with fewer requests. 
Karakoy et al. \cite{2019_Karakoy_ACMmacs}
propose a slicing-based approach
to identify data (memory) accesses 
that can be skipped
to deliver energy/performance gains
within an acceptable error bound. 
The proposed method applies
backward and forward code slicing
to estimate the gains 
from skipping each output data.
Moreover,
the `0' value 
is used for each data access that is not performed. 
The ApproxANN framework \cite{2015_Zhang_DATE},
besides performing approximate computations,
skips memory accesses 
on neural networks
according to the neuron criticality. 
More specifically, 
a theoretical analysis is adopted
to study the impact of neurons on the output accuracy
and 
characterize their criticality. 
The neuron approximation 
under a given QoS constraint
is tuned
by an iterative algorithm,
which applies the approximations
and
updates the criticality of each neuron (it may change due to approximations in other neurons).

\subsection{Approximate Memoization}
The memoization technique stores
results of previous calculations
or pre-computed values
in memory
to use them instead of performing calculations.
Namely,
this memory functions as a look-up table that 
maps 
a set of data identifiers to a set of stored data. 
The current survey focuses on approximate memoization techniques \cite{2011_Chaudhuri_FSE, 2014_Samadi_ASPLOS, 2014_Mishra_WACAS, 2017_Brumar_IPDPS, 2015_Keramidas_WAPCO, 2018_Tziantzioulis_IEEEmicro} relying on software frameworks, compilers and programmer's decisions.
Nevertheless,
it is noted that that there are also approaches \cite{2005_Alvarez_IEEEtc, 2013_Rahimi_IEEEtcasii, 2018_Zhang_IEEEcal, 2019_Liu_ISCA}
requiring hardware modification to support memoization,
as well as hardware-level look-up table
approximation techniques (e.g., \cite{qlut_2017}).
The latter work proposes a quantized look-up table, called qLUT, 
which replaces complex arithmetic functions.
The qLUT table contains precomputed output values
corresponding to
a small input set
that has been created
from the original input data 
using 
their probability distributions. 

Chadhuri et al. \cite{2011_Chaudhuri_FSE} propose an approximate memoization for computations in loops.
Prior to executing an expensive function within a loop,
this technique checks a look-up table
to find
if this computation was previously performed for similar input data.
In this case, 
the cached result is used,
otherwise,
the function is executed 
and the new computation is stored in the look-up table. 
Paraprox \cite{2014_Samadi_ASPLOS} is a software framework
for identifying common patterns in data-parallel programs
and applying tailored approximations to them.
For the Map \& Scatter/Gather patterns,
Paraprox uses memoization rather than performing computations. 
In particular,
it fills a look-up table with pre-computed data,
which are obtained from 
the execution of the Map \& Scatter/Gather function 
for some representative inputs,
and performs runtime look-up table queries
instead of the conventional computations. 

iACT \cite{2014_Mishra_WACAS}
is another approximation framework
that applies runtime memoization among other functionalities.
The programmer uses pragmas
to declare the functions for memoization
and specify the error tolerance percentage.
For each function call-site,
the framework creates a global table
to store pairs of function arguments and output results.
In case the function arguments 
are already stored in the table (within an error bound),
the corresponding output results are returned.
Otherwise,
the function is accurately executed
and the new input--output pairs are stored in the table.
The ATM approach \cite{2017_Brumar_IPDPS}
performs runtime task memoization,
relying on hashing functions to store the task inputs
and an adaptive algorithm to automatically decide whether to use memoization or execute the task.
The programmer needs to 
use pragmas to specify the tasks that are suitable for memoization. 
The authors of 
\cite{2015_Keramidas_WAPCO} 
introduce an approximate memoization
mechanism for GPU fragment shading operations,
which reduces the precision of the input parameters
and performs partial matches.
To identify approximate memoization opportunities,
they characterize various fragment shader instructions
in terms of memoization hits and output accuracy. 
Moreover,
runtime policies are proposed
to tune the precision according to the errors introduced.

Contrary to the aforementioned techniques,
TAF-Memo \cite{2018_Tziantzioulis_IEEEmicro}
is an output-based function memoization technique,
i.e., it memoizes function calls based on their output history.
TAF-Memo checks for temporal locality 
by calculating the relative arithmetic difference of two consecutive
output values from the same function call-site.
In case this difference is below the acceptable error constraint,
memoization is applied by returning the last computed output for the following function calls. 

\subsection{Relaxed Synchronization}

The execution of parallel applications
on multi/many-core systems  
requires time-consuming synchronization
to either access shared data
or satisfy data dependencies. 
Various techniques \cite{2012_Renganarayana_RACES, 2012_Misailovic_RACES, 2013_Misailovic_ACMtecs, 2015_Campanoni_CGO, 2020_Stitt_ACMtecs, 2010_Sreeram_IISWC, 2010_Mengt_IPDPS, 2012_Rinard_RACES}
have been proposed to relax the conventional synchronization requirements that guarantee error-free execution,
delivering speedup 
in exchange for QoS loss. 

The authors of \cite{2012_Renganarayana_RACES} propose the RaC methodology
to systematically relax synchronization,
while always satisfying a programmer-defined QoS constraint. 
Initially,
the programmer specifies the parallel code segments,
and then applies the four-step RaC methodology.
This methodology 
identifies criteria for quantifying the acceptable QoS,
selects the relaxation points,
modifies the code to enable the execution of both the original and relaxed versions,
and selects the suitable relaxation degree (i.e., which instances to relax for each synchronization point). 
Misailovic et al. \cite{2012_Misailovic_RACES}
propose the Dubstep system,
which relaxes the synchronization of parallelized programs
based on a ``find-transform-navigate'' approach.
More specifically,
Dubstep performs a profiling-based analysis of the original program 
to find possible optimizations,
inserts opportunistic synchronization and barriers,
and finally, performs an exploration including   
accuracy, performance and safety analysis 
for the transformed program. 

QuickStep \cite{2013_Misailovic_ACMtecs} is a system for approximately parallelizing sequential programs, 
i.e., without preserving the semantics of the original program,
within statistical accuracy bounds.
Among other transformations,
QuickStep replicates shared objects to eliminate the bottlenecks of synchronized operations on them. 
HELIX-UP \cite{2015_Campanoni_CGO} is another parallelizing compiler
that selectively relaxes strict adherence to program semantics to tackle runtime performance bottlenecks,
involving profiling and user interaction to tune QoS.
The compiler also offers a synchronization-relaxing knob to decrease the inter-core communication overhead
by synchronizing sequential segments with prior iterations. 
More recently, 
the authors of \cite{2020_Stitt_ACMtecs} introduce PANDORA,
which is an approximate parallelizing framework based on symbolic regression machine learning and sampled outputs of the original function.
To avoid timing bottlenecks, such as data movement and synchronization,
and improve parallelism,
PANDORA eliminates loop-carried dependencies using fitness functions and constraints regarding error and performance. 
In \cite{2010_Sreeram_IISWC},
the authors exploit the concept of 
approximate shared value locality
to reduce synchronization conflicts
in programs using optimistic synchronization.
The reduction of conflicts on approximately local variables,
detected for a given similarity constraint,
is achieved through
an arbitration mechanism that
imprecisely shares the values between threads.
The authors of \cite{2010_Mengt_IPDPS} apply aggressive coarse-grained parallelism on recognition \& mining algorithms
by relaxing or even ignoring 
data dependencies between different iterations.
As a result, the timing overheads are reduced in comparison with the conventional parallel implementation,
which also applies parallelization only within the iteration (iterations are executed serially).
Rinard \cite{2012_Rinard_RACES} introduces synchronization-free updates to shared data structures
by eliminating the conventional use of mutual exclusion 
and dropping array elements at the worst scenario.
Moreover,
the same work 
applies relaxed barrier synchronization,
allowing the threads to pass the barrier without stalling to wait for the other threads. 

\subsection{Precision Scaling}
Precision scaling (tuning) refers to the discipline reduction of the numerical precision,
resulting in improved
processing speed
and/or 
memory bandwidth \cite{2020_Cherubin_ACMsrv}.
The state-of-the-art software-level works
\cite{2011_Dinechin_IEEEtc, 2010_Linderman_CGO, 2017_Chiang_POPL, 2017_Darulova_ACMtoplas, 2013_Rubio_SC, 2018_Guo_ISSTA, 2013_Lam_ICS, 2018_Lam_SAGE, 2014_Chiang_PPoPP, 2016_Rubio_ICSE, 2018_Yesil_IEEEmicro, 2015_Tian_GLSVLSI, 2018_Menon_SC, 2020_Brunie_SC, 2019_Laguna_HPC, 2020_Kang_CGO, 2014_Schkufza_PLDI}
address several challenges,
such as 
the scaling degree,
scaling automation,
mixed precision,
and dynamic scaling.

Starting with works based on formal methods to
reduce the precision and examine the errors,
the Gappa tool \cite{2011_Dinechin_IEEEtc}
automates the
study of rounding errors in elementary functions and floating-point calculations using interval arithmetic. 
An extended version of this tool is
Gappa++ \cite{2010_Linderman_CGO},
which provides automated analysis of numerical errors in a wide range of computations,
i.e., 
fixed-point, floating-point, linear and non-linear.
This tool integrates several features,
such as
operation rewriting 
to facilitate the isolation of rounding errors,
and affine arithmetic 
to accurately bound linear calculations with correlated errors.
FPTuner \cite{2017_Chiang_POPL} is a tool that 
performs formal error analysis
based on symbolic Taylor expansions
and quadratically constrained quadratic programming.
It searches for precision allocations
that satisfy constraints such as
the number of
operators at a given precision
and the number of type casts.
Rosa \cite{2017_Darulova_ACMtoplas}
is a source-to-source compiler 
that combines satisfiability modulo theories with interval arithmetic
to bound the round-off errors of the fixed- and floating-point formats.

Several works of the literature employ heuristics and automated search to scale the precision of floating-point programs.
Precimonious \cite{2013_Rubio_SC} 
searches all the program variables in their order of declaration
using the delta-debugging algorithm,
and lowers their precision according to an error constraint specified by the programmer.
In the same context, 
HiFPTuner \cite{2018_Guo_ISSTA}
firstly groups dependent variables 
that may require the same precision,
and then performs a customized hierarchical search. 
Lam et al. \cite{2013_Lam_ICS}
introduce a framework
that employs
the breadth-first search algorithm
to identify code regions that
can tolerate lower precision.
Similarly to this technique,
CRAFT \cite{2018_Lam_SAGE} performs binary searches 
to initially determine the
required program precision,
and then truncate the results of some of the floating-point instructions.
Towards the detection of large floating-point errors,
the authors of \cite{2014_Chiang_PPoPP} propose S$^3$FP.
This tool is  
based on an heuristic-guided search
to find the inputs causing the largest errors.
The Blame Analysis \cite{2016_Rubio_ICSE}  
combines concrete and shadow execution
to generate a blame set for the program instructions,
which contains the minimum precision requirements under a giver error constraint.
This approach can be also used in cooperation with the previous search-based works,
and specifically,
as pre-processing
to reduce the search space. 
Schkufza et al. \cite{2014_Schkufza_PLDI}
treat the scaling of floating-point precision
as a stochastic search problem.
In particular,
they repeatedly 
apply random program transformations
and use a robust search to guarantee the maximum errors.

The concept of dynamic precision scaling,
i.e., the tuning of the precision at runtime with respect to the input data and error sensitivity,   
has been studied in \cite{2018_Yesil_IEEEmicro}. 
The dynamic scaling framework of this work integrates
an offline application profiler, 
a runtime monitor to track workload changes,
and an accuracy controller 
to adjust the precision accordingly.
ApproxMA \cite{2015_Tian_GLSVLSI} dynamically scales the precision of
data memory accesses in algorithms
such as mixture model-based clustering.
This framework integrates a runtime precision controller,
which generates custom bit-widths according to the QoS constraints,
and a memory access controller,
which loads the scaled data from memory.
The custom bit-widths are generated
by analyzing a subset of data and intermediate results
and calculating metrics
regarding the error appearance and the number of tolerable errors.

Mixed floating-point precision has been also studied in high-performance computing workloads.
ADAPT \cite{2018_Menon_SC} uses algorithmic differentiation,
i.e., 
a technique for numerically evaluating the derivative of a function corresponding to the program, 
to estimate the output error of high-performance computing workloads. 
It provides a precision sensitivity 
profile to guide the development of mixed-precision programs.  
In the same context, 
the authors of \cite{2020_Brunie_SC} 
propose an instruction-based search
that explores information 
about the dynamic program behaviour and the temporal locality.

To enable mixed floating-point precision in GPUs,
the authors of \cite{2019_Laguna_HPC} propose
the GPUMixer tool,
which relies on static analysis to
find code regions where precision scaling improves the performance.
Next,
GPUmixer performs a dynamic analysis
involving shadow computations
to examine if the scaled program configurations satisfy the accuracy constraints.
In the same context, 
PreScaler \cite{2020_Kang_CGO} is an automatic framework
that generates precision-scaled OpenCL programs,
considering both kernel execution and data transfer.
Initially, it employs
a system inspector
to collect
information about precision scaling on the target platform,
and an application profiler 
to identify
memory objects with floating-point elements for potential scaling.
This information is exploited by 
a decision maker that 
finds the best scaling configuration using decision tree search on a minimized space.

\subsection{Data Sampling}
Approximate Computing is also exploited in big data analysis,
in an effort to reduce the increased number of computations and storage requirements 
due to the large amount of input data. 
The key idea is to perform computations on a representative data sample
rather than on the entire input dataset.
Therefore, various data sampling methods \cite{2012_Laptev_VLDB, 2015_Goiri_ASPLOS, 2019_Hu_MASCOTS, 2017_Quoc_MIDDL, 2016_Krishnan_WWW, 2017_Quoc_ATC, 2018_Wen_ICDCS, 2013_Agarwal_EuroSys, 2016_Kandula_MOD, 2016_Zhang_VLDB, 2016_Anderson_ICDE, 2019_Park_MOD} 
are examined
to provide real-time processing with error bounds
in applications involving stream analytics, database search, and model training.  

EARL \cite{2012_Laptev_VLDB}
is an extension of Hadoop
(i.e., 
a software framework that provides distributed storage and big data processing on clusters), 
which delivers early results with reliable error bounds.
It applies statistics-based uniform sampling 
from distributed files.
Goiri et al. \cite{2015_Goiri_ASPLOS}
propose the ApproxHadoop framework
to generate approximate MapReduce programs
based on 
task dropping and multi-stage input sampling. 
They also bound the errors
using statistical theories. 
The programmer tunes the approximation
by specifying
either the desired error bound
or the task dropping and input sampling ratios.  
Similarly, 
ApproxSpark \cite{2019_Hu_MASCOTS}
performs sampling at multiple arbitrary points 
of long chains of transformations
to facilitate the aggregation of huge amounts of data.
This framework models 
the clustering information of transformations
as a data provenance tree,
and then computes the approximate aggregate values as well as error thresholds.
Moreover, 
the sampling rates are dynamically selected
according to programmer-specified error thresholds.

Sampling methods have been also examined
in stream analytics.
StreamApprox \cite{2017_Quoc_MIDDL}
is an approximate stream analytics system
that supports both batched and pipelined stream processing.
It employs two sampling techniques,
i.e., stratified and reservoir sampling,
to approximate the outputs with rigorous error bounds.
IncApprox \cite{2016_Krishnan_WWW} 
combines approximate and incremental computations to 
provide stream analytics with bounded error.
This system executes a stratified sampling algorithm
that selects data for which the results have been memoized from previous runs,
and adjusts the computations
to produce an incrementally updated output.
On the other hand,
PrivApprox \cite{2017_Quoc_ATC} 
combines sampling and randomized response 
to provide both approximate computations and privacy guarantee.
This system integrates a query execution interface
that enables the systematic exploration of the trade-off
between accuracy and query budget.
ApproxIoT \cite{2018_Wen_ICDCS}
facilitates approximate stream analytics
at the edge
by combining stratified reservoir sampling
and hierarchical processing.

A variety of sampling methods have been employed
in approximate query processing systems for databases.
BlinkDB \cite{2013_Agarwal_EuroSys}
performs approximate distributed query processing,
supporting SQL-based aggregation queries 
with time and error constraints.
It creates stratified samples based on past queries,
and uses an heuristic-based profiler to dynamically
select the sample that meets the query's constraints. 
Another system applying approximate big-data queries 
is Quickr \cite{2016_Kandula_MOD},
which integrates operators 
sampling multiple join inputs
into a query optimizer,
and then searches for an appropriate 
sampled query plan.
Sapprox \cite{2016_Zhang_VLDB} is a distribution-aware system
that employs the occurrences of sub-datasets
to drive the online sampling.
In particular, 
the exponential number of sub-datasets is reduced 
to a linear one using a probabilistic map, 
and then,
cluster sampling with unequal probability theory 
is applied for sub-dataset sampling. 
Sapprox also determines the optimal sampling unit size 
in relation with approximation costs and accuracy.

Numerous works of the literature
use data sampling to decrease the increased computational cost of model training
in machine learning applications.
Zombie \cite{2016_Anderson_ICDE}
is a two-stage system
that trains approximate models based on clustering and active learning.
The first stage applies offline indexing
to organize the dataset into index groups of similar elements.
Subsequently,
the stage of online querying 
uses the index groups that are likely to output useful features
to creates the training subset of data. 
BlinkML \cite{2019_Park_MOD} approximately trains a model on a small sample,
while providing accuracy guarantees.
The sample is obtained through uniform random sampling,
however, in case of very large datasets, 
a memory-efficient algorithm is employed.

\subsection{Approximate Programming Languages}
The high-level approximation of software programs has been examined
through approximate programming languages,
i.e., language extensions 
that allow the programmer to 
systematically declare approximate code regions, variables, loops, and functions,
insert randomness in the program,  
and/or specify error constraints. 
The literature involves numerous works \cite{2011_Ansel_CGO, 2007_Sorber_SenSys, 2010_Baek_PLDI, 2015_Boston_OOPSLA, 2012_Carbin_PLDI, 2013_Carbin_OOPSLA, 2014_Misailovic_OOPSLA, 2011_Liu_ASPLOS, 2015_Achour_OOPSLA, 2011_Sampson_PLDI, 2015_Park_FSE, 2014_Park_GIT, 2008_Goodman_UAI, 2014_Mansinghka_CoRR, 2016_Tolpin_IFL, 2014_Bornholt_ASPLOS, 2014_Sampson_PLDI, 2019_Fernando_ACMpapl, 2019_Joshi_ICSE}
that enable approximate procedural, object-oriented, and probabilistic programming. 

Ansel et al.
\cite{2011_Ansel_CGO}
introduce a set of PetaBricks language extensions that allow the programmer to write code of variable accuracy.
These extensions expose
the performance--accuracy trade-off 
to the compiler,
which automatically searches the algorithmic space 
to tune the program according to the programmer's accuracy constraints.
Eon \cite{2007_Sorber_SenSys} is a programming language 
that allows the programmer to annotate program flows (paths) with different energy states.
The Eon runtime system predicts the workload and energy of the system,
and then adjusts the execution of flows according to the programmer's declarations and the energy constraints. 
In the same context, 
Baek and Chilimbi \cite{2010_Baek_PLDI} propose Green,
which is a two-phase programming framework
providing language extensions 
to approximate expensive functions and loops.
The programmer uses pragma-like annotations to 
specify approximate variants of functions.
In the calibration phase,
Green builds a model to quantify the QoS loss and the performance/energy gains.   
This model is then used in the operational phase
to generate an approximate program satisfying the programmer's QoS constraint. 
DECAF \cite{2015_Boston_OOPSLA} is a type-based approximate programming language
that allows the programmer
to specify the correctness probability 
for some of the program variables. 
The DECAF type system also integrates
solver-aided type inference
to automatically tune the type of the rest variables,
code specialization,
and dynamic typing.
Flikker \cite{2011_Liu_ASPLOS} provides language annotations 
to mark the program variables
and partition the data into critical and non-critical regions 
(the latter are stored in unreliable memories).
Topaz \cite{2015_Achour_OOPSLA}
is a task-based language
that maps tasks onto approximate hardware
and uses an outlier detector 
to find and re-execute the computations
producing unacceptable results.

In \cite{2012_Carbin_PLDI},
the authors introduce
language constructs for generating approximate programs
and proof rules for verifying the acceptability properties.
Rely \cite{2013_Carbin_OOPSLA} is an imperative language that allows the programmer 
to introduce quantitative reliability specifications
for generating programs with data stored in approximate memory and inexact arithmetic/logical operations.
Chisel \cite{2014_Misailovic_OOPSLA}
automates the selection of Rely's approximations
while satisfying the programmer-defined reliability and accuracy constraints.
To solve this optimization problem, 
Chisel employs an integer programming solver. 
All these works include safety analysis and program verification for sequential programs.
In contrast, 
Parallely \cite{2019_Fernando_ACMpapl} is a programming language for approximating parallel programs
through canonical sequentialization, 
i.e., 
a verification method that generates sequential programs capturing the semantics of parallel programs.

Targeting approximations in Java programs,
the authors of \cite{2011_Sampson_PLDI}
propose EnerJ, i.e., a language extension providing type qualifiers to specify data that can be approximately stored or computed.
EnerJ guarantees isolation of the approximate computations.
FlexJava \cite{2015_Park_FSE} offers another
set of language extensions 
to annotate approximate programs.
Using an approximation safety analysis,
FlexJava
automates the approximation of data and operations 
while ensuring safety guarantees. 
ExpAX \cite{2014_Park_GIT}
allows the programmer to explicitly specify error expectations for a subset of Java.
Based on an approximation safety analysis, 
it identifies operations that are candidate for approximation,
and then,
a heuristic-based framework approximates those that statistically satisfy the error expectations.

Significant research has also been conducted on probabilistic programming languages.
Church \cite{2008_Goodman_UAI} is a probabilistic language that inserts randomness on a deterministic function subset using stochastic functions.
The Church semantics are defined in terms of 
evaluation histories and conditional distributions on the latter.
Similarly, 
Venture \cite{2014_Mansinghka_CoRR} is another language
that enables the specification of probabilistic models and inference problems.
The Anglican \cite{2016_Tolpin_IFL} language and runtime system
provides probabilistic evaluation model and functional representations, 
e.g., distributions and sequences of random variables. 

Uncertain\raisebox{0.8pt}{$\scriptstyle <$}T\raisebox{0.8pt}{$\scriptstyle >$} \cite{2014_Bornholt_ASPLOS}
is a language abstraction
that manipulates data as probability distributions.
More specifically, 
random variables are declared as ``uncertain''
and a Bayesian network for representing computations is build,
where nodes correspond to the variables
and edges correspond to conditional variable dependencies.
The Uncertain\raisebox{0.8pt}{$\scriptstyle <$}T\raisebox{0.8pt}{$\scriptstyle >$} runtime system 
performs hypothesis tests and sampling to evaluate the network.
In the same context, 
Sampson et al. \cite{2014_Sampson_PLDI}
use probabilistic assertions on random variables.
Their tool,
called MayHap,
performs probabilistic evaluation by
statically building a Bayesian representation network based on the input distribution
and
dynamically interpreting it via sampling.
In the same context,
AxProf \cite{2019_Joshi_ICSE}
is a profiling-based framework for 
analyzing randomized approximate programs.
The programmer specifies probabilistic predicates for the output, 
i.e., regarding the expectation of the output value
and/or the probability that the output satisfies a condition, 
and AxProf generates
approximate programs based on statistical tests.

%% file: TBs/tb3_sw_classification.tex
\begin{table*}[!t]
\renewcommand{\arraystretch}{1.05}
\centering
\footnotesize
\setlength{\tabcolsep}{5pt}
\caption{Classification of software approximation techniques.}
\vspace{-8pt}
\begin{tabular}{l l}
\hline  
\makecell[l]{\textbf{SW Approximation Class}} & \makecell[c]{\textbf{References}}\\
\hline
\hline
Loop Perforation &  \cite{2009_Hoffmann_MIT, 2010_Misailovic_ICSE, 2011_Sidiroglou_FCE, 2015_Shi_IEEEcal, 2017_Omar_ICCD, 2018_Li_ICS, 2020_Baharvand_IEEEtetc, 2015_Tan_ASP-DAC, 2018_Kanduri_DAC} \\
\hline
Computation Skipping & \cite{2010_Byna_GPGPU, 2015_Raha_DATE, 2015_Vassiliadis_PPoPP, 2016_Vassiliadis_CGO, 2006_Rinard_ICS, 2007_Rinard_OOPSLA, 2017_Lin_ISCAS, 2018_Akhlaghi_ISCA} \\
\hline
Memory Access Skipping & \cite{2014_Miguel_MICRO, 2016_Yazdanbakhsh_ACMtaco, 2018_Kislal_ELSEclss, 2013_Samadi_MICRO, 2019_Karakoy_ACMmacs, 2015_Zhang_DATE} \\
\hline
Approximate Memoization & \cite{2011_Chaudhuri_FSE, 2014_Samadi_ASPLOS, 2014_Mishra_WACAS, 2017_Brumar_IPDPS, 2015_Keramidas_WAPCO, 2018_Tziantzioulis_IEEEmicro, 2005_Alvarez_IEEEtc, 2013_Rahimi_IEEEtcasii, 2018_Zhang_IEEEcal, 2019_Liu_ISCA, qlut_2017} \\
\hline
Relaxed Synchronization & \cite{2012_Renganarayana_RACES, 2012_Misailovic_RACES, 2013_Misailovic_ACMtecs, 2015_Campanoni_CGO, 2020_Stitt_ACMtecs, 2010_Sreeram_IISWC, 2010_Mengt_IPDPS, 2012_Rinard_RACES} \\
\hline
Precision Scaling & \cite{2011_Dinechin_IEEEtc, 2010_Linderman_CGO, 2017_Chiang_POPL, 2017_Darulova_ACMtoplas, 2013_Rubio_SC, 2018_Guo_ISSTA, 2013_Lam_ICS, 2018_Lam_SAGE, 2014_Chiang_PPoPP, 2016_Rubio_ICSE, 2018_Yesil_IEEEmicro, 2015_Tian_GLSVLSI, 2018_Menon_SC, 2020_Brunie_SC, 2019_Laguna_HPC, 2020_Kang_CGO, 2014_Schkufza_PLDI} \\
\hline
Data Sampling & \cite{2012_Laptev_VLDB, 2015_Goiri_ASPLOS, 2019_Hu_MASCOTS, 2017_Quoc_MIDDL, 2016_Krishnan_WWW, 2017_Quoc_ATC, 2018_Wen_ICDCS, 2013_Agarwal_EuroSys, 2016_Kandula_MOD, 2016_Zhang_VLDB, 2016_Anderson_ICDE, 2019_Park_MOD} \\
\hline
Approx. Programming Languages &   \cite{2011_Ansel_CGO, 2007_Sorber_SenSys, 2010_Baek_PLDI, 2015_Boston_OOPSLA, 2012_Carbin_PLDI, 2013_Carbin_OOPSLA, 2014_Misailovic_OOPSLA, 2011_Liu_ASPLOS, 2015_Achour_OOPSLA, 2011_Sampson_PLDI, 2015_Park_FSE, 2014_Park_GIT, 2008_Goodman_UAI, 2014_Mansinghka_CoRR, 2016_Tolpin_IFL, 2014_Bornholt_ASPLOS, 2014_Sampson_PLDI, 2019_Fernando_ACMpapl, 2019_Joshi_ICSE} \\
\hline \hline
\end{tabular}
\label{tb_sw}
\end{table*}

%% file: 5_hardware.tex
\section{Hardware Approximation Techniques}\label{sec:hw_lvl}

\begin{figure}[!t]
\centering
\includegraphics[width=0.85\textwidth]{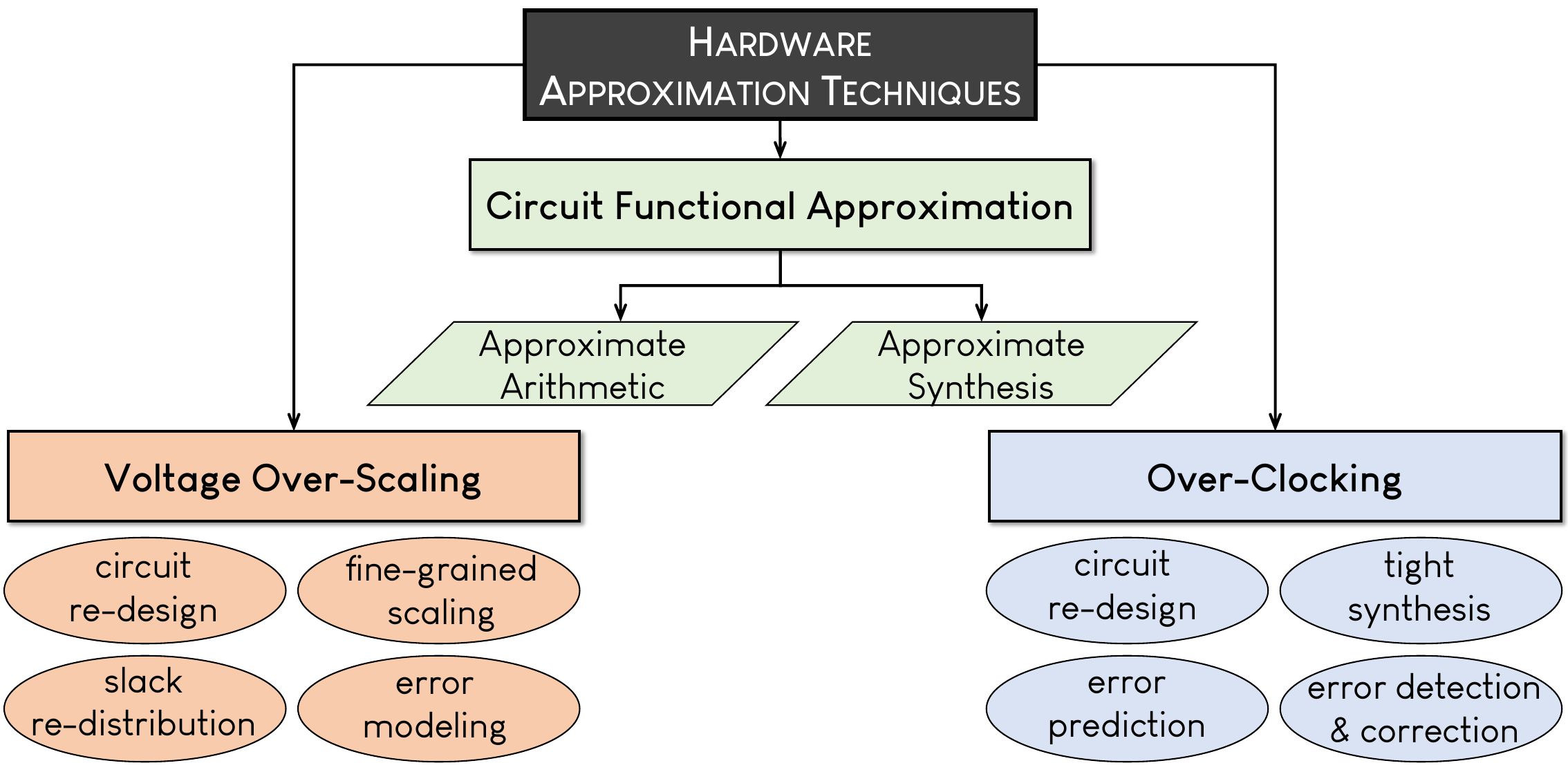}
    \vspace{-6pt}
    \caption{Classification of HW approximation techniques in 3 classes: \emph{Circuit Functional Approximation}, \emph{Voltage Over-Scaling} and \emph{Over-Clocking}.}
    \label{fg_hw}
\end{figure}

This section classifies and introduces the hardware approximation techniques,
which are applied at the lower level of the design abstraction hierarchy.
These techniques aim to 
improve the area, power consumption, and performance 
of the circuits
i.e., the basic building blocks of accelerators, processors, and computing platforms.
The hardware approximation techniques can be categorized into three classes:
(i) \emph{Circuit Functional Approximation (CFA)},
(ii) \emph{Voltage Over-Scaling (VOS)},
and (iii) \emph{Over-Clocking (OC)}.
In approximate hardware, two types of errors are distinguished: 
the functional errors (produced by CFA)
and the timing errors (produced by VOS and OC).
Fig.~\ref{fg_hw} illustrates the
hardware approximation techniques,
including a further taxonomy
to sub-classes.

The remainder of this section presents state-of-the-art works, organized according to the proposed classification of
Table \ref{tb_hw}. 
It is noted that,
even though some works may belong to more than one sub-class,
they are assigned  
to their prevalent one  
and their relevant features are highlighted. 

\input{TBs/tb4_hw_classification}

\subsection{Circuit Functional Approximation}
Circuit functional approximation
modifies the original accurate design
by reducing its circuit complexity at logic level.
Typical CFA approaches include:  
(i) the modification of the circuit's truth table,
(ii) the use of an approximate version of the initial hardware algorithm,
(iii) the use of small inexact components as building blocks,
and
(iv) approximate circuit synthesis.
The main target of CFA is the arithmetic circuits \cite{2020_Jiang_IEEE},
as they constitute the key processing units of processors and accelerators, 
and thus,
they inherently affect their power efficiency and performance. 
The literature provides several open-source libraries of approximate arithmetic circuits,
such as ApproxAdderLib \cite{2015_Shafique_DAC}, EvoApprox8b \cite{2017_Mrazek_DATE} and SMApproxLib \cite{2018_Ullah_DAC}.
This survey focuses on approximate adders, multipliers, and dividers.
However, 
it is noted that that numerous works design and evaluate other approximate arithmetic operations, 
such as
circuits for 
Multiplication-and-Accumulation (MAC) \cite{2019_Chen_IEEEtc, 2019_Gillani_IEEEa}, 
square root \cite{2019_Jiang_IEEEtc},
squaring \cite{2020_Manikanta_IEEEtvlsi},
square-accumulate \cite{2018_Gillani_IEEEa},
and Coordinate Rotation Digital Computer (CORDIC) \cite{2017_Chen_IEEEtscl}.
The literature also includes 
automated methods
for generating approximate circuits,
which are presented in the context of approximate logic synthesis. 
Moreover, 
there are works applying functional approximations based on the inputs such as \cite{date21}, where the authors configure and assign different approximation operators for a given data flow program based on the input workload. 
It is also important to mention that methodologies for digital hardware design based on approximate arithmetic circuits and formats have been proposed \cite{leon_fpl}. 

\subsubsection{Approximate Adders}
Significant research has been conducted on the design of approximate area- and power-efficient adders.
The approximation techniques for inexact adders involve:
(i) \emph{use of approximate full adder cells} \cite{2013_Gupta_IEEEtcad, 2013_Yang_NANO, 2018_Pashaeifar_IEEEtvlsi, 2018_Dalloo_IEEEtvlsi}
and
(ii) \emph{segmentation and carry prediction} \cite{2013_Kim_ICCAD, 2015_Hu_DATE, 2012_Kahng_DAC, 2013_Ye_ICCAD, 2015_Shafique_DAC, 2018_Akbari_IEEEtcasii, 2018_Xu_IEEEtvlsi, 2020_Ebrahimi_IEEEtcasii}.
Next, representative state-of-the-art works for approximate adders are presented.

The IMPACT adders
are based on inexact full adders cells,
which are approximated
at the transistor level to deliver up to 45\%  
area reduction \cite{2013_Gupta_IEEEtcad}.
Another transistor-level cell approximation is proposed in \cite{2013_Yang_NANO},
where 
the AXA 4-transistor XOR/XNOR-based adders are implemented,
delivering up to 31\% gain in dynamic power consumption.
Moreover, 
in \cite{2018_Pashaeifar_IEEEtvlsi}, 
approximate reverse carry propagate full adders 
are used to build the hybrid RCPA adders.
Targeting higher level approximations, 
the OLOCA adder
splits the addition into accurate and approximate segments \cite{2018_Dalloo_IEEEtvlsi},
and for the latter,
it employs OR gates for the most-significant bit additions
and outputs constant `1' for the least-significant ones.

To reduce the worst-case carry propagation delay,
Kim et al. \cite{2013_Kim_ICCAD}
propose a carry prediction scheme
leveraging the less-significant input bits,
which
is 2.4$\times$ faster than the conventional ripple-carry adder.
Similarly,
Hu et al. \cite{2015_Hu_DATE}
introduce 
a carry speculating method
to segment the carry chain
in their design,
which also performs error and sign correction.
Compared to the accurate adder,
the proposed design
is 4.3$\times$ faster 
and saves 47\% power. 

The quality constraint of applications may vary during runtime,
thus,
research efforts have also focused on 
designing dynamically configurable adders that 
can tune their accuracy. 
In \cite{2012_Kahng_DAC},
the authors propose an accuracy-configurable adder,
called ACA, 
which consists of several sub-adders and an error detection \& correction module.
The accuracy is controlled at runtime,
while operation in accurate mode is also supported. 
Another dynamically configurable adder,
called GDA,
is proposed in \cite{2013_Ye_ICCAD},
where multiplexers select the carry input either from the 
previous sub-adder or the carry prediction unit,
providing a more graceful accuracy degradation.
In the same direction,
the GeAr adder \cite{2015_Shafique_DAC} 
employs multiple sub-adders
of equal length 
to variable approximation modes. 
This architecture 
supports accurate mode via a
configurable error correction unit. 

Akbari et al. \cite{2018_Akbari_IEEEtcasii}
introduce the RAP-CLA adder,
which
splits the conventional 
carry look-ahead scheme
into two segments,
i.e., the approximate part and the augmenting part,
supporting approximate and accurate mode.
When operating at the approximate mode,
the augmenting part is power-gated to reduce power consumption.
Another carry-prediction-based approach supporting both modes
is the SARA design \cite{2018_Xu_IEEEtvlsi}.
This adder uses carry ripple sub-adders,
and the carry prediction does not require a dedicated circuitry.
Finally, the BSCA adder,
which is based on a block-based carry speculative approach \cite{2020_Ebrahimi_IEEEtcasii},
integrates an error recovery unit
and
non-overlapped blocks consisting of 
a sub-adder, a carry prediction unit, and a selection unit.

\subsubsection{Approximate Multipliers}
The multiplication circuits have attracted significant interest from the research community.
The literature includes a plethora of inexact multipliers that 
can be categorized according to the prevailing approximation techniques:
(i) \emph{truncation and rounding} \cite{2015_Hashemi_ICCAD, 2017_Zendegani_IEEEtvlsi, 2018_Leon_IEEEmicro, 2021_Leon_ACMtecs, 2019_Vahdat_IEEEtvlsi, 2019_Leon_DAC, 2020_Frustaci_IEEEtcasii}, 
(ii) \emph{approximate radix encodings} \cite{2017_Liu_IEEEtc, 2019_Venkatachalam_IEEEtc, 2016_Jiang_IEEEtc, 2020_Waris_IEEEtcasii, 2021_Waris_IEEEtcasii, 2018_Leon_IEEEtvlsi, 2022_Zhu_IEEEtcasii},
(iii) \emph{use of approximate compressors}     \cite{2015_Momeni_IEEEtc, 2017_Akbari_IEEEtvlsi, 2019_Sabetzadeh_IEEEtcasi, 2018_Esposito_IEEEtcasi,  2020_Strollo_IEEEtcasi},
and (iv) \emph{logarithmic approximation} \cite{2018_Liu_IEEEtcasi, 2020_Saadat_DATE, 2021_Ansari_IEEEtc, 2021_Pilipovic_IEEEtcasi}.
Subsequently,
the state-of-the-art works from each category are introduced. 

Starting with the rounding and truncation techniques,
the DRUM multiplier \cite{2015_Hashemi_ICCAD}
dynamically reduces the input bit-width, 
based on the leading `1' bits,
to achieve 60\% power gain in exchange for mean relative error of 1.47\%.
Zendegani et al.
propose
the RoBa multiplier \cite{2017_Zendegani_IEEEtvlsi},
which 
rounds the operands to the nearest exponent-of-two 
and performs a shift-based multiplication in segments.
In \cite{2018_Leon_IEEEmicro},
the PR approximate multiplier
perforates partial products
and applies rounding to the remaining ones,
delivering up to 69\% energy gains.
The same approximation technique is integrated 
in the mantissa multiplication of floating-point numbers
to create the AFMU multiplier \cite{2021_Leon_ACMtecs}.
Vahdat et al.
propose
the TOSAM multiplier \cite{2019_Vahdat_IEEEtvlsi} that 
truncates the input operands 
according to their leading `1' bit.
To decrease the error,
the truncated values are rounded to the nearest odd number.  
In \cite{2019_Leon_DAC},
different rounding, perforation and encoding schemes are combined to
extract the most energy-efficient multiplication circuits. 
Finally,
Frustaci et al. \cite{2020_Frustaci_IEEEtcasii}
implement an alternative dynamic truncation with correction,
along with an efficient mapping for the remaining partial product bits.

Next,
multipliers that 
generate their partial products
based on approximate radix encodings are presented. 
Liu et al. \cite{2017_Liu_IEEEtc}
modify the Karnaugh map of the radix-4 encoding
to create approximate encoders for generating the least-significant partial product bits.
A similar approach is followed in \cite{2019_Venkatachalam_IEEEtc},
where approximate radix-4 partial product generators are designed.
Jiang et al. \cite{2016_Jiang_IEEEtc}
use an approximate adder to generate the $\pm$3$\times$ multiplicand term in the radix-8 multiplier.
In \cite{2020_Waris_IEEEtcasii},
the authors propose 
a hybrid low-radix encoding that 
encodes the most-significant bits
with the 
accurate radix-4 encoding
and the least-significant bits with the an approximate radix-8 encoding.
A similar approach is used in \cite{2021_Waris_IEEEtcasii},
where the authors propose 
approximate radix-8 multipliers for
FPGA-based design.  
Targeting to high-order radix, 
the authors of \cite{2018_Leon_IEEEtvlsi}
propose 
the hybrid high-radix encoding,
which applies both the
accurate radix-4 and approximate radix-2$^k$ encodings. 
Correspondingly,
in \cite{2022_Zhu_IEEEtcasii},
a radix-256 encoder is proposed 
for approximate multiplication circuits.

Several works employ approximate compressors for 
the partial product accumulation.
Momeni et al. \cite{2015_Momeni_IEEEtc}
modify the truth table of the accurate
4:2 compressor 
to create two simplified designs
and use them in the Dadda multiplier.
The authors in \cite{2017_Akbari_IEEEtvlsi}
design 4:2 compressors,
again for Dadda multipliers,
which can switch
between accurate and approximate mode at runtime,
consuming 68\% lower power.
In \cite{2019_Sabetzadeh_IEEEtcasi},
an approximate 4:2 compressor
is implemented in FinFET
based on 
a three-input majority gate,
and then it is used in the Dadda architecture along truncation.
Esposito et al. \cite{2018_Esposito_IEEEtcasi}
introduce a new family of approximate compressors 
and assign them to 
each column of the partial product matrix
according to their allocation algorithm.
Another interesting work
is the design of approximate compressors for multipliers using
the stacking circuit concept \cite{2020_Strollo_IEEEtcasi}.

Regarding the approximate logarithmic multipliers,
Liu et al. \cite{2018_Liu_IEEEtcasi}
employ
a truncated binary-logarithm converter
and inexact adders for the mantissa addition
to design the ALM family of multipliers.
The logarithmic-based REALM multiplier \cite{2020_Saadat_DATE}
partitions the 
power-of-two intervals
of the input operands
into segments,
and
determines an error compensation factor 
for each one.
The ILM multiplier \cite{2021_Ansari_IEEEtc}
differentiates from the conventional design,
as it 
rounds the input operands to their 
nearest power-of-two
using a nearest `1' bit detector.
Pilipovic et al. \cite{2021_Pilipovic_IEEEtcasi}
propose
a two-stage trimming 
logarithmic multiplier,
which firstly, 
reduces the bit-width of
the input operands,
and then, 
the bit-width of the
mantissas.

\subsubsection{Approximate Dividers}
The division circuits have received less attention
than adders and multipliers.
Nevertheless,
the literature provides numerous works 
aiming to reduce the large critical paths of the conventional dividers.
The approximation techniques for division circuits can be categorized as follows:
(i) \emph{bit-width scaling} \cite{2016_Hashemi_DAC, 2019_Jiang_IEEEtc},
(ii) \emph{use of approximate adder/subtractor cells} \cite{2015_Chen_GLSVLSI, 2016_Chen_IEEEtc, 2018_Chen_IEEEtmscs, 2020_Adams_IEEEtc},
and (iii) \emph{simplification of computations} \cite{2016_Zendegani_DATE, 2017_Vahdat_DATE, 2019_Imani_Date,  2018_Liu_ARITH, 2019_Saadat_DAC}.

The first class of approximation techniques
uses exact dividers with reduced bit-width.
The approximate divider of \cite{2016_Hashemi_DAC}
dynamically selects the most relevant bits
from the input operands
and 
performs accurate division at lower bit-width,
providing up to 70\% power gains in exchange for 3\% average error.
The design makes use of leading `1' bit detectors,
priority encoders,
multiplexers,
subtractor
and barrel shifter.
Similarly, 
the AAXD divider of \cite{2019_Jiang_IEEEtc}  
detects the leading `1' bits
and 
uses a pruning
scheme to extract the bits
that will be given as input to the divider. 
Additionally, 
the design integrates an error correction unit 
to form the final output.

Regarding the second class of approximation techniques,
Chen et al. \cite{2015_Chen_GLSVLSI}
perform the subtraction of the non-restoring array divider with inexact subtractor circuits employing 
pass transistor logic.
For their divider,
called AXDnr,  
the authors examine different schemes
regarding which subtractions of the division array to approximate. 
Similarly,
in the AXDr divider of \cite{2016_Chen_IEEEtc},
some of the subtractions of the restoring array divider are performed with inexact subtractor circuits. 
The use of inexact cells has also been examined
in
the high-radix SRT divider \cite{2018_Chen_IEEEtmscs}.
In this divider,
called HR-AXD,
the inexact cell is 
a signed-digit adder that 
is employed according to different replacement schemes,
along with cell truncation and error compensation.
Adams et al. \cite{2020_Adams_IEEEtc}
introduce two approximate division architectures,
called AXRD-M1 and AXRD-M2, 
which deliver up to 46\% area and 57\% power gains, respectively, 
compared to the exact restoring divider.
The first design replaces 
some of the restoring divider cells with
inexact ones of simplified logic,
while the second one 
involves the elimination of some rows of the divider.

Targeting to perform the division with alternative simplified computations,  
the SEERAD divider \cite{2016_Zendegani_DATE}
rounds the divisor to a specific form 
based on the leading `1' bit position,
and thus,
the division is transformed to shift-\&-add multiplication.
In the same context, 
Vahdat et al. \cite{2017_Vahdat_DATE}
propose the TruncApp divider   
that multiplies  
the truncated dividend
with the approximate inverse
divisor.
Targeting to model the division operation, 
the CADE divider of \cite{2019_Imani_Date} 
performs the floating-point division 
by subtracting the input mantissas.
To compensate a large error
(estimated by analyzing the 
most-significant input bits),  
a pre-computed value is retrieved from memory.
In \cite{2018_Liu_ARITH},
the proposed AXHD divider
approximates the least-significant computations of the division
using an non-iterative logarithmic approach that 
is based on leading `1' bit detection 
and subtraction of the logarithmic
mantissas. 
Finally,
Saadat et al. \cite{2019_Saadat_DAC}
propose approximate integer and floating-point dividers with near-zero error bias,
called INZeD and FaNZeD, respectively,
by combining an error correction method with the classical approximate logarithmic divider.

\subsubsection{Approximate Synthesis}
An automated approach to
generate inexact circuits
is the approximate logic synthesis. 
This method
provides increased approximation diversity,
i.e., it generates multiple approximate circuit variants, 
without relying on 
the manual approximation inserted by the designer,
such as in the case of the aforementioned arithmetic approximations.
Another benefit of approximate synthesis 
is that 
several techniques 
generate the approximate variant
that leads  
to the maximum hardware gains
for a given approximation/error constraint. 
The state-of-the-art techniques
can be categorized as follows \cite{2020_Scarabottolo_IEEE}:
(i) \emph{structural netlist transformation}
\cite{2017_Schlachter_IEEEtvlsi, 2018_Scarabottolo_DATE, 2013_Venkataramani_DATE, 2017_Liu_ICCAD, 2021_Castro_IEEEtcasii}, 
(ii) \emph{Boolean rewriting} 
\cite{2012_Venkataramani_DAC, 2014_Ranjan_DATE, 2013_Miao_ICCAD, 2018_Hashemi_DAC}, 
(iii) \emph{high-level approximate description} \cite{2015_Yazdanbakhsh_DATE, 2014_Nepal_DATE, 2017_Lee_DATE, 2019_Nepal_IEEEtetc, 2020_Castro_ICCAD},
and (iv) \emph{evolutionary synthesis} \cite{2013_Sekanina_ICES, 2015_Vasicek_IEEEtec, 2017_Mrazek_DATE, 2016_Vasicek_FPL, 2019_Vasicek_DATE}.

Several works of the literature employ
a direct acyclic graph
to represent the circuit netlist,
where each node corresponds to a gate. 
In this context,
the GLP technique \cite{2017_Schlachter_IEEEtvlsi}
prunes nodes with an iterative greedy approach
according to their impact on
the final output and their toggle activity.
In contrast,
the CC framework \cite{2018_Scarabottolo_DATE}
performs an exhaustive exploration of
all possible
node subsets that can be pruned
without surpassing the error constraint.
Venkataramani et al. \cite{2013_Venkataramani_DATE}
propose SASIMI,
which is based on a greedy heuristic
to find signal pairs assuming the same value
and substitute one with the other.
This automatic synthesis framework 
guarantees that the user-defined quality constraint is satisfied,
and generates accuracy-configurable circuits.
To apply stochastic netlist transformation,
the SCALS framework \cite{2017_Liu_ICCAD}
maps an initial gate-level network to the targeted technology (standard-cell or FPGA),
and then iteratively extracts sets of sub-netlists
and
inserts random approximations in them.  
These sub-netlists are evaluated
using statistical hypothesis testing. 
Castro-Codinez et al. \cite{2021_Castro_IEEEtcasii}
propose the AxLS framework,
which converts the Verilog netlist to XML format
and then applies typical transformation techniques, 
e.g., gate pruning,
with respect to an error threshold.

The second category 
includes techniques
that apply approximations in a formal Boolean representation of the circuit 
before it is synthesized.
The SALSA approach \cite{2012_Venkataramani_DAC} 
encodes the error constraints
into a quality logic function,
which compares the outputs of the accurate and approximate circuits.
Towards logic simplification,
SALSA computes
the ``observability don't cares''
for each output of the approximate circuit,
i.e.,
the set of input values for which 
the output is insensitive.
In the same direction,
but for sequential circuits,
Ranjan et al. introduce
ASLAN \cite{2014_Ranjan_DATE}.
This framework 
generates several approximate variants
of the 
combinational blocks,
and then identifies the best
approximations for the entire sequential circuit
based on a gradient-descent approach.
Miao et al. \cite{2013_Miao_ICCAD} 
use a two-phase Boolean minimization algorithm
to address the problem of approximate synthesis.
The first phase solves the problem under a given constraint for error magnitude, 
and the second phase iteratively finds a solution that also
satisfies the error frequency constraint.
In an iterative manner, 
the BLASYS methodology \cite{2018_Hashemi_DAC}
partitions the circuit into smaller circuits,
and for each one,
it generates an approximate truth table 
based on 
Boolean matrix factorization.
The approximate sub-circuits are synthesized 
and the trade-off between error and power/area efficiency
for the entire circuit is evaluated. 

Regarding approximations introduced at the hardware description level,
Yazdanbakhsh et al. \cite{2015_Yazdanbakhsh_DATE}
propose the Axilog language annotations,
which provide syntax and semantics for approximate design and reuse in Verilog.
Axilog allows the designer 
to partition the design into 
accurate and approximate segments.
ABACUS \cite{2014_Nepal_DATE} is another interesting work that 
parses
the behavioral Verilog description of the design
to create its abstract syntax tree.
Next, 
a set of diverse transformations
is applied to the tree
to create approximate variants, 
which are then
written in Verilog.
An expanded version of ABACUS is introduced in
\cite{2019_Nepal_IEEEtetc},
where sorting-based evolutionary algorithms are employed for design space exploration.
Moreover, 
the new ABACUS version focuses on approximations
in critical paths
to facilitate 
the reduction of the supply voltage.
Lee et al. \cite{2017_Lee_DATE}  
generate approximate designs in Verilog 
from C accurate descriptions.
The proposed framework
computes 
data statistics and mobility information for
the given design,
and employs an heuristic solver
for optimizing the energy--quality trade-off.
Targeting to high-level synthesis,
the AxHLS approach
\cite{2020_Castro_ICCAD}
performs a design space exploration 
based on analytical models 
to identify the best arithmetic approximations
for a given error constraint.
Starting from a C description,
AxHLS adopts
scheduling and binding
operations
to apply the approximations provided by the exploration
and generate the Verilog code.

The fourth class of techniques for automated synthesis of approximate circuits is based on evolutionary algorithms,
i.e., heuristic-based search algorithms that treat circuit approximation
as multi-objective optimization problem and generate a set of solutions.
In this context,
Sekanina et al. \cite{2013_Sekanina_ICES}
use Cartesian genetic programming
to minimize the error in adders
considering the number of logic gates as constraint. 
This approach is extended in \cite{2015_Vasicek_IEEEtec},
where approximate multipliers and median filters
are evolved through randomly seeded Cartesian genetic programming. 
Based on the same utilities, 
the authors of \cite{2017_Mrazek_DATE} propose 
the EvoApprox8b library of approximate adders and multipliers.
This library is generated by examining various trade-offs between accuracy and hardware efficiency,
and offers different approximation variants and circuit architectures. 
In \cite{2016_Vasicek_FPL},
a search-based technique for evolutionary circuit synthesis for FPGAs is proposed.
In particular, 
this approach represents the circuit as a directed acyclic graph,
and 
re-synthesizes approximate configurations based on Cartesian genetic programming.
Vasicek et al. \cite{2019_Vasicek_DATE}
adjust the approximation degree
with respect to the significance of the inputs.
To do so, they
adopt a weighted error metric to determine the significance of each input vector and
use Cartesian genetic programming to minimize the circuit’s area while satisfying a
threshold. 

\subsection{Voltage Over-Scaling}
Voltage over-scaling aims to 
reduce the circuit's supply voltage
below its nominal value,
while keeping 
the clock frequency constant. 
The circuit operation 
at a lower voltage value 
produces timing errors
due to the failure of the critical paths
to meet the delay constraints.
Nevertheless, 
considering that power consumption
depends on the voltage value,
VOS techniques are continously examined in the literature.
An exploration and quantification of the benefits and overheads of VOS is presented in \cite{2010_Kurdahi_IEEEtvlsi}.
Research involving VOS can be classified in the following categories:
(i) \emph{slack re-distribution} \cite{2010_Kahng_ASP-DAC},
(ii) \emph{circuit re-design and architecture modification} \cite{2011_Mohapatra_DATE, 2013_Chen_IEEEtvlsi, 2018_Zhang_DAC},
(iii) \emph{fine-grained scaling} \cite{2019_Pandey_DAC, 2020_Wang_IEEEtc, 2019_Zervakis_IEEEtcasii},
and 
(iv) \emph{error modeling} \cite{2010_Liu_IEEEtvlsi, 2012_Jeon_IEEEtcasii, 2017_Ragavan_DATE, 2020_Jiao_IEEEtcad,  2018_Zervakis_IEEEtvlsi}.

Kahng et al.
\cite{2010_Kahng_ASP-DAC} 
shift the timing slack
of the frequently executed near-critical paths
through slack redistribution,
and thus,
reduce the minimum voltage
at which the error rate remains acceptable.
The proposed technique is based on post-layout cell resizing
to deliver the switching activity-aware slack redistribution. 
More specifically,
a heuristic
finds the voltage
satisfying the desired error rate,
and then increases the transistor width 
of the cells
to optimize the frequently executed paths.

In \cite{2011_Mohapatra_DATE},
the authors
optimize building blocks
for more graceful degradation under VOS,
using two techniques,
i.e., 
dynamic segmentation \& error compensation
and
delay budgeting of chained datapath.
The first technique
bit-slices the datapath of the adder 
and employs a multi-cycle error correction circuitry that tracks the carries.
The second technique 
adds transparent latches 
between chained arithmetic units
to distribute the clock period.
To facilitate VOS,
Chen et al. \cite{2013_Chen_IEEEtvlsi}
build their designs
on
the residue number system,
which
provides shorter critical paths
than conventional arithmetic.
They also
employ the reduced precision redundancy scheme
to eliminate the timing errors. 
Another interesting work is Thundervolt \cite{2018_Zhang_DAC}, 
which provides error recovery
in the MAC units
of systolic arrays. 
To detect timing errors,
Thundervolt
employs Razor shadow flip-flops. 
In case an error occurs in a MAC,
a multiplexer 
forwards the previous MAC's accurate partial sum
(stored in the Razor flip-flop)
to the next MAC.

Targeting fine-grained VOS solutions,
i.e., 
the use of different voltages
across the same circuit architecture, 
Pandey et al. 
propose GreenTPU \cite{2019_Pandey_DAC}.
This technique    
stores 
input sequences producing timing errors in MACs.
As a result,
when such an input sequence pattern
is identified,
the voltage of the MAC
is scaled accordingly 
to prevent timing errors. 
In the same context,
the authors of \cite{2020_Wang_IEEEtc}
propose NN-APP.
This framework 
analyzes the error propagation 
in neural networks
to model 
the impact of VOS on accuracy.
Based on this analysis,
as well as an error resilience study for the neurons,
NN-APP uses 
a voltage clustering method
to assign the same voltage
to neurons with similar error resilience.
Another fine-grained VOS approach is proposed in
\cite{2019_Zervakis_IEEEtcasii}.
This framework 
provides voltage heterogeneity
by using
a greedy algorithm to   
solve the optimization problem
of grouping and assigning the voltage of arithmetic units
to different islands. 

The analysis of errors
in circuits under VOS is considered a key factor,
as it guides the aggressiveness of voltage scaling
towards the acceptable error margins.
In \cite{2010_Liu_IEEEtvlsi},
an analytical method to study the errors in voltage over-scaled arithmetic circuits
is proposed.
Similarly, 
the authors of 
\cite{2012_Jeon_IEEEtcasii}
introduce a probabilistic approach to
model the errors
of the critical paths.
In the same category, 
works relying on simulations to analyze the errors of VOS can be included. 
Ragavan et al.
\cite{2017_Ragavan_DATE}
characterize 
arithmetic circuits in terms of energy efficiency and errors
using 
transistor-level SPICE simulation for various voltages. 
Based on this characterization,
they propose a statistical model 
to simulate the behavior of arithmetic operations in VOS systems.
By exploiting machine learning methods,
Jiao et al. \cite{2020_Jiao_IEEEtcad}
propose
LEVAX
to model voltage over-scaled functional units.
This input-aware model
is trained on data from gate-level simulations 
to
predict the timing error rate
for each output bit.
To provide accurate VOS-aware gate-level simulation,
Zervakis et al. propose VOSsim \cite{2018_Zervakis_IEEEtvlsi}.
This framework
performs an offline 
characterization of the flip-flop
for timing violations,
and 
calculates the cell delays
for the targeted voltage,
enabling gate-level simulation under VOS.

\subsection{Over-Clocking}
Over-clocking (or frequency over-scaling)
aims to operate the circuit/system at higher clock frequencies
than those that respect the critical paths.
As a result,
timing errors are induced in exchange for increased performance.
A trade-off analysis between accuracy and performance 
when over-clocking FPGA-based designs
is presented in \cite{2013_Shi_FCCM}.
In the same work, 
the authors show that OC 
outperforms the traditional bit truncation for the same error constraint. 
The analysis of the current survey considers that the state-of-the-art works of the domain focus on the following directions: 
(i) \emph{tight synthesis} \cite{2018_Alan_IEEEtcad}, 
(ii) \emph{circuit re-design and architecture modification} \cite{2013_Ramasubramanian_DAC, 2014_Shi_DAC, 2017_Wang_IEEEtvlsi}, 
(iii) \emph{error detection \& correction} \cite{2014_Choudhury_IEEEtc, 2016_Ragavan_ISVLSI, 2017_Li_IEEEtvlsi},
and 
(iv) \emph{error prediction} \cite{2012_Roy_DAC, 2015_Constantin_DATE, 2016_Jiao_ICCD, 2017_Jiao_DATE, tevot, devot}.

The first approach towards
the reduction of timing errors caused by OC 
optimizes the critical paths of the design.
In this context, 
the SlackHammer framework \cite{2018_Alan_IEEEtcad}
synthesizes circuits with tight delay constraints
to 
reduce the number of near-critical paths,
and thus, 
decrease the probability of timing errors when frequency is over-scaled. 
At first, 
SlackHammer isolates the paths
and identifies potential delay optimizations. 
Based on the isolated path analysis,
the framework performs an iterative synthesis
with tighter constraints for the 
primary outputs with negative slack. 

The second class of techniques 
aims at modifying the conventional circuit architecture
to facilitate frequency OC and increase the resilience to timing errors.
The retiming technique \cite{2013_Ramasubramanian_DAC}
re-defines the boundaries of combinational logic
by moving the flip-flops backward or forward between the stages.
Based on this circuit optimization, 
the synthesis is relaxed
by ignoring the 
paths that are bottleneck to minimum period retiming. 
Targeting different circuit architectures, 
Shi et al. \cite{2014_Shi_DAC}
adopt an alternative arithmetic,
called Online, 
and show that online-based circuits are
more resilient to the timing errors of OC
than circuits with traditional arithmetic.
The modification of the initial neural network model
to provide resilience in timing errors
has also attracted research interest. 
In this direction, 
Wang et al. \cite{2017_Wang_IEEEtvlsi} 
propose an iterative reclocking-and-retraining framework for  
operating neural network circuits at higher frequencies
under a given accuracy constraint.
The clock frequency is gradually increased, and the network's weights are updated through back-propagation training
until to find the maximum frequency for which
the timing errors are mitigated
and the accuracy constraint is satisfied.

Several works propose circuits for timing error detection \& correction,
enabling the use of over-clocking.
These techniques either 
improve the frequency value of the first failure,
i.e., the first timing error,
or 
reduce the probability of timing errors.  
TIMBER \cite{2014_Choudhury_IEEEtc}
masks timing errors
by borrowing time from successive pipeline stages.
According to this approach,
the use of 
discrete time-borrowing flip-flops
and continuous time-borrowing latches
slows down the appearance of timing errors
with respect to the frequency scaling.
Ragavan et al. \cite{2016_Ragavan_ISVLSI}
detect and correct timing errors by
employing a dynamic speculation window 
on the double-sampling scheme.
This technique 
adds an extra register,
called shadow and clocked by a second ``delayed'' clock, 
at the end of the pipelined path
to sample the output data at two different time instances.
This approach also
uses an online slack measurement to adaptively
over-clock the design.
The TEAI approach \cite{2017_Li_IEEEtvlsi}
is based on 
the locality of the timing errors
in software-level instructions,
i.e., the tendency of specific instructions to produce timing errors.
TEAI identifies these instructions at runtime,
and sends error alarms
to hardware,
which is equipped with error detection \& correction circuits.

Significant research has also been conducted on predicting the timing errors in advance, 
allowing to over-scale the frequency 
according to the acceptable error margins. 
In \cite{2012_Roy_DAC},
the authors introduce an
instruction-level error prediction system
for pipelined micro-processors,
which stalls the pipeline when
critical instructions are detected.
Their method is based on
gate-level simulations to
find the critical paths that are 
sensitized during the program execution.
Similarly,
Constantin et al.
\cite{2015_Constantin_DATE}
obtain the maximum delays for 
each arithmetic instruction 
through gate-level simulations,
and 
dynamically exploit timing margins
to apply frequency over-scaling.

In addition to 
instruction-level prediction models,
there are numerous works 
that build models 
based on machine learning 
and simulations of functional units.
A representative work of this approach 
is WILD \cite{2016_Jiao_ICCD},
which builds a 
workload-dependent prediction model
using logistic regression. 
In the same direction,
SLoT \cite{2017_Jiao_DATE} is a supervised learning model that 
predicts timing errors
based on the inputs and the clock frequency.
At first, 
SLoT performs gate-level simulation
to extract timing class labels, 
i.e., ``timing error'' or ``no timing error'', 
for different inputs and frequencies.
These classes are then used, 
along with features extracted from random data pre-processing, 
to train the error prediction model.
Towards the same approach, 
TEVoT \cite{tevot} uses machine learning to build a timing error prediction model that can predict the timing errors under different clock speeds and operating conditions, which are used to estimate the output quality of error-tolerant applications (e.g., image processing). 
DEVoT \cite{devot} is an extension of TEVoT that formulates the timing error prediction as a circuit dynamic delay prediction problem, saving significant prediction resources. 

%% file: TBs/tb4_hw_classification.tex
\begin{table*}[!t]
\renewcommand{\arraystretch}{1.05}
\centering
\footnotesize
\setlength{\tabcolsep}{15pt}
\caption{Classification of hardware approximation techniques.}
\vspace{-8pt}
\begin{tabular}{l l}
\hline  
\makecell[l]{\textbf{HW Approximation Class}} & \makecell[c]{\textbf{Technique/Approach}}\\
\hline
\hline
\multirow{2}{*}{Adder Approximation} & Use of Approximate Full Adder Cells \cite{2013_Gupta_IEEEtcad, 2013_Yang_NANO, 2018_Pashaeifar_IEEEtvlsi, 2018_Dalloo_IEEEtvlsi} \\
& Segmentation and Carry Prediction \cite{2013_Kim_ICCAD, 2015_Hu_DATE, 2012_Kahng_DAC, 2013_Ye_ICCAD, 2015_Shafique_DAC, 2018_Akbari_IEEEtcasii, 2018_Xu_IEEEtvlsi, 2020_Ebrahimi_IEEEtcasii} \\
\hline 
\multirow{4}{*}{Multiplier Approximation} & Truncation and Rounding \cite{2015_Hashemi_ICCAD, 2017_Zendegani_IEEEtvlsi, 2018_Leon_IEEEmicro, 2021_Leon_ACMtecs, 2019_Vahdat_IEEEtvlsi, 2019_Leon_DAC, 2020_Frustaci_IEEEtcasii} \\
& Approximate Radix Encodings \cite{2017_Liu_IEEEtc, 2019_Venkatachalam_IEEEtc, 2016_Jiang_IEEEtc, 2020_Waris_IEEEtcasii, 2021_Waris_IEEEtcasii, 2018_Leon_IEEEtvlsi, 2022_Zhu_IEEEtcasii} \\
& Use of Approximate Compressors     \cite{2015_Momeni_IEEEtc, 2017_Akbari_IEEEtvlsi, 2019_Sabetzadeh_IEEEtcasi, 2018_Esposito_IEEEtcasi,  2020_Strollo_IEEEtcasi} \\
& Logarithmic Approximation \cite{2018_Liu_IEEEtcasi, 2020_Saadat_DATE, 2021_Ansari_IEEEtc, 2021_Pilipovic_IEEEtcasi} \\
\hline
\multirow{3}{*}{Divider Approximation} & Bit-width Scaling \cite{2016_Hashemi_DAC, 2019_Jiang_IEEEtc} \\
& Use of Approximate Adder/Subtractor Cells \cite{2015_Chen_GLSVLSI, 2016_Chen_IEEEtc, 2018_Chen_IEEEtmscs, 2020_Adams_IEEEtc} \\
& Simplification of Computations \cite{2016_Zendegani_DATE, 2017_Vahdat_DATE, 2019_Imani_Date,  2018_Liu_ARITH, 2019_Saadat_DAC} \\
\hline
\multirow{4}{*}{Approximate Synthesis} & Structural Netlist Transformation
\cite{2018_Scarabottolo_DATE, 2013_Venkataramani_DATE, 2017_Liu_ICCAD, 2021_Castro_IEEEtcasii} \\
& Boolean Rewriting
\cite{2012_Venkataramani_DAC, 2014_Ranjan_DATE, 2013_Miao_ICCAD, 2018_Hashemi_DAC} \\
& High-Level Approximate Description \cite{2015_Yazdanbakhsh_DATE, 2014_Nepal_DATE, 2017_Lee_DATE, 2019_Nepal_IEEEtetc, 2020_Castro_ICCAD} \\
& Evolutionary Synthesis \cite{2013_Sekanina_ICES, 2015_Vasicek_IEEEtec, 2017_Mrazek_DATE, 2016_Vasicek_FPL, 2019_Vasicek_DATE} \\
\hline
\multirow{4}{*}{Voltage Over-Scaling} & Slack Re-distribution \cite{2010_Kahng_ASP-DAC} \\
& Circuit Re-design and Architecture Modification \cite{2011_Mohapatra_DATE, 2013_Chen_IEEEtvlsi, 2018_Zhang_DAC} \\
& Fine-Grained Scaling \cite{2019_Pandey_DAC, 2020_Wang_IEEEtc, 2019_Zervakis_IEEEtcasii} \\
& Error Modeling \cite{2010_Liu_IEEEtvlsi, 2012_Jeon_IEEEtcasii, 2017_Ragavan_DATE, 2020_Jiao_IEEEtcad,  2018_Zervakis_IEEEtvlsi} \\
\hline
\multirow{4}{*}{Over-Clocking} & Tight Synthesis \cite{2018_Alan_IEEEtcad} \\
& Circuit Re-design and Architecture Modification \cite{2013_Ramasubramanian_DAC, 2014_Shi_DAC, 2017_Wang_IEEEtvlsi} \\
& Error Detection \& Correction \cite{2014_Choudhury_IEEEtc, 2016_Ragavan_ISVLSI, 2017_Li_IEEEtvlsi} \\
& Error Prediction \cite{2012_Roy_DAC, 2015_Constantin_DATE, 2016_Jiao_ICCD, 2017_Jiao_DATE, tevot, devot} \\
\hline \hline
\end{tabular}
\label{tb_hw}
\end{table*}

%% file: 6_comparison.tex
\section{Comparative Quantitative Analysis of Approximation Techniques}\label{sec:comp}

This section reports a quantitative analysis for the software and hardware approximation classes. 
Due to their very large volume, diversity and differentiation,  
direct intra- and cross-layer 
comparisons are not performed.
However, significant outcomes can be extracted
for each class 
(e.g., type and size of workloads, acceptable accuracy loss, targeted resource gains). 
It is also noted that additional comparisons are reported in Part II of the survey \cite{mysurvey_pt2}, where an application-driven analysis of case studies is presented, involving both software and hardware techniques.  

\subsection{Software Approximation Techniques}
Table \ref{tb_sw_comp} reports 
remarkable software-level works from each approximation class, 
along with key numerical results for resource gains and errors. 
Based on the literature's review, each software approximation technique is favored in specific workloads (e.g., precision scaling in high-performance floating-point programs and data sampling in large queries),
even though there are also more general techniques (e.g., loop perforation).  
Most works evaluate the approximations for various levels of quality degradation, 
e.g., 2\%, 5\%, and 10\%, 
with the latter being widely considered as 
the largest acceptable threshold for the workloads of Table \ref{tb_sw_comp}. 
Another significant outcome is that the combination
of approximation knobs, such as in the case of approximate programming languages, 
delivers more resource gains 
than the application of a single approximation technique.
For example, in comparison with approximate memoization,
the approximate programming language of Table \ref{tb_sw_comp}
provides 8\%--28\% more energy gain for workloads of similar type and size. 

\input{TBs/tb5_sw_comparison}

\begin{figure}[!b]
    \centering
    \includegraphics{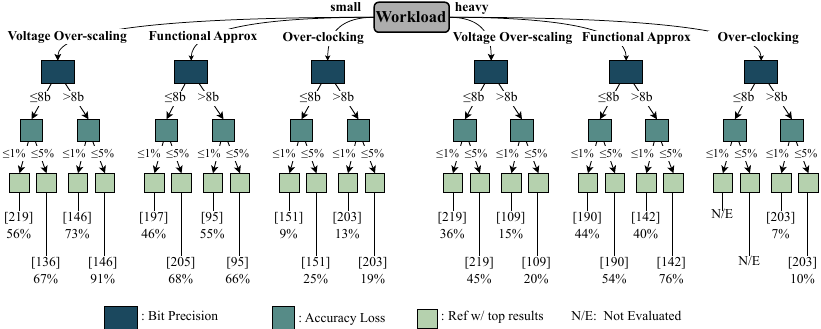}
    \caption{Classification and comparative analysis of the most remarkable hardware approximation works with respect to their workload complexity, baseline bit precision and an accuracy loss threshold (1\% and 5\%). The leafs present the corresponding work along with their top results in energy reduction.}
    \label{fig:comp_tree}
\end{figure}

\subsection{Hardware Approximation Techniques}
The respective results for all the hardware approximation techniques analyzed in the previous sections are summarized in Fig.~\ref{fig:comp_tree}.
To generate Fig.~\ref{fig:comp_tree}, we identified the most commonly used workloads among each hardware approximation family and categorized them into two arbitrary levels: \textit{small} and \textit{heavy}. 
Workloads requiring more than 50M operations (such as in deep neural networks) are classified as heavy, while those below this threshold are considered small.
Then, we created a decision tree that helps the reader to identify the works having remarkable results with respect to the complexity of the workload, the baseline bit precision and the desired accuracy loss constraint. For the accuracy loss, 
two thresholds are considered, i.e., small accuracy loss (less than 1\%) and moderate accuracy loss (less than 5\%). For only few reported works, we define low accuracy loss as an MRED of less than 5\%, while medium accuracy loss falls between 5\% and 15\%.
The leaves represent the works that achieve the highest energy reduction in each case.
Overall, it seems that both Voltage Over-scaling and Circuit Functional Approximation appear among the top energy reduction results, outperforming Over-clocking.
It is worth noting that Over-clocking shows a significant lack of energy efficiency compared to other techniques when processing heavy workloads. On the other hand, approximation techniques in small workloads can achieve substantial energy savings (e.g., 68\%, 91\%), even when using reduced precision and considering low accuracy loss thresholds. 
However, the high difference between small and heavy workloads highlights the need for continued research into hardware approximation techniques. 

%% file: TBs/tb5_sw_comparison.tex
\begin{table*}[!t]
\renewcommand{\arraystretch}{1.05}
\centering
\footnotesize
\setlength{\tabcolsep}{4pt}
\caption{Quantitative analysis of software approximation techniques.}
\vspace{-8pt}
\begin{tabular}{c c c c c}
\hline  
\makecell[c]{\textbf{Approximation Class}} & 
\makecell[c]{\textbf{Workloads}} & 
\makecell[c]{\textbf{Resource Gains}} &
\makecell[c]{\textbf{Accuracy Metric}} &
\makecell[c]{\textbf{Ref.}}
\\
\hline
\hline
\multirow{2}{*}{Loop Perforation} & \emph{BlackScholes}, \emph{KMeans} &  32\%--36\% energy & rel.err=3.5\%, PSNR=36dB & \cite{2020_Baharvand_IEEEtetc}\\
                 & \emph{X264}, \emph{Streamcluster} &  2$\times$--8.5$\times$ speedup & error=5\%,10\% & \cite{2018_Li_ICS}\\
\hline
\multirow{2}{*}{Computation Skipping} &  \emph{GoogLeNet}, \emph{VGGNet} & 2.2$\times$--1.9$\times$ speedup &  class.accur.loss$\le$3\% & \cite{2018_Akhlaghi_ISCA}\\
& \emph{MPEG}, \emph{KNN} & $\sim$2$\times$ energy & qual.degrad=2.5\%,5\% & \cite{2015_Raha_DATE} \\
\hline
\multirow{2}{*}{Memory Access Skipping} &  \emph{FluidAnimate}, \emph{BodyTrack} & 6\%--28\% speedup & error<10\% & \cite{2014_Miguel_MICRO} \\
& \emph{Gaussian}, \emph{MatMul} & 1.45$\times$--2.4$\times$ speedup & qual.degrad=10\% & \cite{2016_Yazdanbakhsh_ACMtaco} \\
\hline
\multirow{2}{*}{Approximate Memoization} 
& \emph{BoxMuller}, \emph{GammaCorr} & 1.5$\times$--3.2$\times$ speedup & qual.degrad=10\% & \cite{2014_Samadi_ASPLOS} \\
& \emph{BodyTrack}, \emph{Sobel} & 22\% energy & qual.degrad=4\%--10\% & \cite{2014_Mishra_WACAS} \\
\hline
\multirow{2}{*}{Relaxed Synchronization} 
& \emph{Graph500}, \emph{KMeans} & 3$\times$--15$\times$ speedup & qual.degrad=0 &  \cite{2012_Renganarayana_RACES} \\
& \emph{Barnes-Hut}, \emph{VolRender} & 6$\times$ speedup &  qual.degrad=0 &  \cite{2012_Misailovic_RACES}\\
\hline
\multirow{2}{*}{Precision Scaling} 
& \emph{Lulesh}, \emph{JetEngine} & 1.2$\times$--1.4$\times$ speedup & error=10$^{-13}$--10$^{-11}$ &  \cite{2018_Menon_SC} \\
& \emph{Bessel}, \emph{FFT} & 11.5$\times$--43.4$\times$ speedup & error=10$^{-10}$--10$^{-4}$ & \cite{2018_Guo_ISSTA} \\
\hline
\multirow{2}{*}{Data Sampling} 
& \emph{WikiLength} & 21\% speedup & error=0.34\% & \cite{2015_Goiri_ASPLOS} \\
& \emph{Conviva}, \emph{TPC-H} & 10$\times$--100$\times$ speedup & accur.loss=2\%-10\% &  \cite{2013_Agarwal_EuroSys} \\
\hline
\multirow{2}{*}{Approx. Programm. Languages}
& \emph{PageRank}, \emph{Sobel} & 1.1$\times$--1.7$\times$ speedup & e2e.error=10$^{-8}$--10$^{-16}$ &  \cite{2019_Fernando_ACMpapl} \\
& \emph{MonteCarlo}, \emph{RayTracer} &  30\%--50\% energy & error<0.2 & \cite{2011_Sampson_PLDI} \\
\hline \hline
\end{tabular}
\label{tb_sw_comp}
\end{table*}

%% file: 7_conclusion.tex
\section{Conclusion}\label{sec:conc}
This article presented Part I of a comprehensive survey on Approximate Computing, focusing on key aspects of this novel design paradigm (motivation, terminology, and principles) and reviewing the state-of-the-art software and hardware approximation techniques. The review and classification was performed in both coarse-grained and fine-grained manners: each software/hardware-level technique was assigned to a higher-level approximation class (e.g., precision scaling, voltage over-scaling), as well as to a lower-level class with respect to its technical/implementation details (e.g., radix encoding, error prediction). Finally, a quantitative analysis of the approximation techniques was reported, involving the most commonly used workloads and key numerical results. Part II of the survey reviews the state-of-the-art software \& hardware application-specific approximation techniques and architecture-level approximations in processors and memories. It also presents the application spectrum of Approximate Computing,
including an analysis of use cases with remarkable results per technique and application domain,
as well as well-established benchmark suites and error metrics for Approximate Computing. 

%% file: 8_acknowledgement.tex
\section*{Acknowledgement}

This research is partially supported by ASPIRE, the technology program management pillar of Abu Dhabi’s Advanced Technology Research Council (ATRC), via the ASPIRE Awards for Research Excellence.